\documentclass[fleqn,usenatbib,useAMS]{mnras}

\usepackage{graphicx}	% Including figure files
\usepackage{amsmath}	% Advanced maths commands
\usepackage{amssymb}	% Extra maths symbols
\usepackage{multicol}        % Multi-column entries in tables
\usepackage{bm}		% Bold maths symbols, including upright Greek
\usepackage{pdflscape}	% Landscape pages
\usepackage[normalem]{ulem}
\usepackage{tikz}
\usetikzlibrary{shapes.geometric, arrows}
\usepackage{multirow}
\usepackage{utfsym}

\tikzstyle{input} = [rectangle, minimum width=1cm, minimum height=1cm, text centered, draw=black, fill=blue!10]

\tikzstyle{output} = [rectangle, minimum width=1cm, minimum height=1cm, text centered, draw=black, fill=white]

\tikzstyle{process} = [rectangle, minimum width=1cm, minimum height=1cm, text centered, draw=black, fill=orange!30]

\tikzstyle{decision} = [rectangle, minimum width=1cm, minimum height=1cm, text centered, draw=black, fill=green!30]

\tikzstyle{arrow} = [thick,->,>=stealth]

\usepackage[T1]{fontenc}
\usepackage{ae,aecompl}

\usepackage{newtxtext,newtxmath}
\title[Unified analysis]{Analysis of Unified Galaxy Power Spectrum Multipole Measurements}

\author[J. Donald-McCann et al.]{
Jamie Donald-McCann,$^{1}$\thanks{E-mail: jamie.donald-mccann@port.ac.uk}
Rafaela Gsponer,$^{1}$
Ruiyang Zhao,$^{2,3,1}$
Kazuya Koyama,$^{1}$
Florian Beutler$^{4}$
\\
$^{1}$Institute of Cosmology \& Gravitation, University of Portsmouth, Dennis Sciama Building, Portsmouth, PO1 3FX, UK\\
$^{4}$Institute for Astronomy, University of Edinburgh, Royal Observatory, Blackford Hill, Edinburgh EH9 3HJ, UK\\
$^{2}$National Astronomy Observatories, Chinese Academy of Science, Beijing, 100101, P.R.China\\
$^{3}$University of Chinese Academy of Sciences, Beijing 100049, P.R.China
}

\date{Accepted XXX. Received YYY; in original form ZZZ}

\pubyear{2023}

\begin{document}
\label{firstpage}
\pagerange{\pageref{firstpage}--\pageref{lastpage}}
\maketitle

\begin{abstract}
We present a series of full-shape analyses of galaxy power spectrum multipole measurements from the 6dFGS, BOSS, and eBOSS galaxy surveys. We use an emulated effective field theory of large-scale structure (EFTofLSS) model to conduct these analyses. We exploit the accelerated prediction speed of the neural-network-based emulator to explore various analysis setups for our cosmological inference pipeline. Via a set of mock full-shape analyses of synthetic power spectrum multipoles, designed to approximate measurements from the surveys above, we demonstrate that the use of alternative priors on nuisance parameters and restricted model complexity reduces many of the biases previously observed in marginalised cosmological constraints coming from EFTofLSS analyses. The alternative priors take the form of a Jeffreys prior; a non-informative prior that can mitigate against biases induced by marginalising over poorly constrained nuisance parameters. When performing a joint analysis of all synthetic multipoles, we see an improvement in the level of agreement between the marginalised $\ln{\left(10^{10}A_s\right)}$ constraints and the truth; from $\sim2.0\sigma$ to $\sim0.42\sigma$. Using our pipeline to analyse the measured multipoles, we find an improvement in the level of agreement with cosmic microwave background (CMB) results; from $\sim2.4\sigma$ to $\sim0.5\sigma$. Therefore, we conclude that the spectroscopic galaxy survey datasets listed above are consistent with constraints obtained from the CMB.
\end{abstract}

\begin{keywords}
large-scale structure of the Universe -- methods: data analysis -- cosmology: cosmological parameters 
\end{keywords}

%%%%%%%%%%%%%%%%%%%%%%%%%%%%%%%%%%%%%%%%%%%%%%%%%%

%%%%%%%%%%%%%%%%% BODY OF PAPER %%%%%%%%%%%%%%%%%%

% The MNRAS class isn't designed to include a table of contents, but for this document one is useful.
% I therefore have to do some kludging to make it work without masses of blank space.

\section{Introduction}
\label{sec:intro}
Conducting \textit{full-shape} analyses of galaxy clustering statistics \citep{satpathy_clustering_2017,kobayashi_full-shape_2021,chen_new_2022,lange_constraints_2023}, such as the power spectrum, is becoming a standard approach to complement analyses that focus of specific features like the baryon acoustic oscillations (BAO). To run one of these full-shape analyses, we require a theoretical model that allows us to make a prediction for the clustering statistic of interest for a given set of cosmological parameters $\bmath{\theta}$. There are two possible routes here: 1.) use a simulation-based model, 2.) use an analytical model. A simulation-based model will likely be more accurate on small, nonlinear, scales. Comparisons of dark matter only N-body simulation codes have shown agreement in predictions of the dark matter power spectrum for scales $k \lesssim 1 \ h\ \mathrm{Mpc}^{-1}$ \citep{schneider_matter_2016,grove_desi_2022}. However, developing a simulation-based model requires many simulations with different sets of cosmological parameters sampling from the parameter space of interest. These suites of simulations \citep[e.g.][]{heitmann_coyote_2010,maksimova_abacussummit_2021} require huge computational cost to produce, and this cost can prohibit the use of such models. An analytic model may be less accurate on nonlinear scales \citep{foreman_precision_2016, alkhanishvili_reach_2022}, but using such a model will incur a significantly lower computational cost.\\

One such analytical model that is gaining in popularity when conducting full-shape analyses is the \textit{effective field theory of large-scale structure} \citep[EFTofLSS;][]{baumann_cosmological_2012,carrasco_effective_2012,senatore_bias_2015,de_la_bella_matter_2017,philcox_combining_2020,ivanov_effective_2022,mergulhao_effective_2023,moretti_modified_2023}. This perturbation-theory based model maps predictions for the dark matter clustering to that of galaxies via a series of nuisance parameters $\bmath{\phi}$, that are marginalised over when putting constraints on the cosmological parameters $\bmath{\theta}$. Two popular examples of EFTofLSS code implementations are \textsc{PyBird} \citep{damico_limits_2021} and \textsc{CLASS-PT} \citep{chudaykin_non-linear_2020}. Predictions for the galaxy power spectrum multipoles can be made with \textsc{PyBird} in $\mathcal{O}(1\ \mathrm{s})$\footnote{This is a processor dependant statement. In \citet{donald-mccann_matryoshka_2022} the prediction speed was reported as $1.01\ \mathrm{s}\pm 13.1\ \mathrm{ms}$. Based on 100 predictions made on a laptop with an Intel i5 2.50 GHz dual-core processor with four threads and 8 GB of RAM. Table 1 of \citep{chudaykin_non-linear_2020} reports prediction speeds from \textsc{CLASS-PT}. In default mode, the performance appears similar to \textsc{PyBird}.}. This is significantly faster than a numerical simulation, but running an MCMC with \textsc{PyBird} still requires a non-negligible amount of computational resources. This cost can limit the exploration of the analysis setup when using this model to carry out parameter inference.\\

The idea of \textit{emulation} to reduce computational cost is being used more and more frequently for cosmological inference problems and is now used to accelerate inference pipelines that are based on analytic theory models \citep{albers_cosmicnet_2019,arico_accelerating_2022,derose_neural_2022,mancini_cosmopower_2022,gunther_cosmicnet_2022,eggemeier_comet_2022,gunther_uncertainty-aware_2023,nygaard_connect_2023} as well as those with simulation-based models \citep{heitmann_cosmic_2006,agarwal_pkann_2014,nishimichi_dark_2019,euclid_collaboration_euclid_2021,storey-fisher_aemulus_2022}. These \text{emulators} consist of nonlinear interpolators that are fitted to (or trained with) a set of input and output pairs $\{\bmath{\theta}, Y(\bmath{\theta})\}$, with $Y(\bmath{\theta})$ being the function of interest. The nonlinear interpolation scheme generally takes the form of a machine learning algorithm like a Gaussian process or neural network (NN). In \citet{donald-mccann_matryoshka_2022}, the NN-based \textsc{EFTEMU} was added to the \textsc{matryoshka} suite of emulators \citep{donald-mccann_span_2022}. The \textsc{EFTEMU} was developed to reduce the cost of EFTofLSS model evaluations and increased the prediction speed of the galaxy power spectrum multipoles by over three orders of magnitude. This increase in prediction speed opens up the opportunity to test more analysis setup choices when using the EFTofLSS model.\\

In this paper, we exploit the increased prediction speed from the \textsc{EFTEMU} to perform full-shape analyses of galaxy power spectrum multipole measurements from several completed galaxy surveys. We also examine how the analysis setup impacts the inferred cosmology. Through a series of mock full-shape analyses, we validate our cosmological inference pipeline. We then demonstrate that using alternative priors and more restrictive sets of nuisance parameters can alleviate some of the biases in the inferred cosmological parameters that can be seen when conducting full-shape analyses with the EFTofLSS. We find that using these alternative priors can alleviate some of the slight tensions in the marginalised cosmological parameter constraints when comparing with results from cosmic microwave background (CMB) analyses. The paper is organised as follows. In Section \ref{sec:data}, we introduce the galaxy surveys considered for this work, along with the multipole measurements used. In Section \ref{sec:model}, we further introduce the EFTofLSS and discuss any changes made to the \textsc{EFTEMU} for this work. In Section \ref{sec:mock_analyses}, we present a series of mock analyses designed to test our inference pipeline. In Section \ref{sec:main_results}, we present results from the analysis of the multipole measurements introduced in Section \ref{sec:data}. We conclude in Section \ref{sec:conclusions}.

\section{Data}
\label{sec:data}
There have now been several large-scale spectroscopic redshift surveys that have run to completion; combining to provide detailed maps of the universe covering a wide redshift range. For this work, we focus on three surveys that cover distinct redshift ranges: the \textit{6dF galaxy survey} \citep[6dFGS,][]{jones_6df_2004,jones_6df_2009}, the \textit{baryon oscillation spectroscopic survey } \citep[BOSS,][]{dawson_baryon_2013,alam_clustering_2017}, and the \textit{extended baryon oscillation spectroscopic survey} \citep[eBOSS,][]{dawson_sdss-iv_2016,eboss_collaboration_completed_2021}. The redshift catalogues from each of these surveys are now publicly available such that galaxy clustering measurements can be made for each of them. \citet{beutler_unified_2021} presents measurements of the power spectrum multipoles from each of these surveys, along with wide-angle and window function matrices. These matrices allow wide-angle effects and the survey window function to be included in theory predictions of the galaxy power spectrum multipoles via two simple matrix multiplications. All measurements have 40 $k$-bins over the range $0 < k < 0.4 \ h\ \mathrm{Mpc}^{-1}$. The BOSS and eBOSS samples are split into subsamples for the northern and southern galactic cap (NGC and SGC) and, in the case of BOSS, two redshift bins (BOSSz1 and BOSSz3). This results in seven sets of multipoles with four effective redshifts $z_\mathrm{eff}=[0.096, 0.38, 0.61, 1.52]$. We refer the reader to Table 1 in \citet{beutler_unified_2021} for more details about each sample.\\

\subsection{Mocks}

\label{subsubsec:pybird_mocks}

When exploring analysis setups, we need to examine if a particular setup leads to more or less bias in the inferred cosmological parameters than another. Mock multipoles were published alongside the measurements in \citet{beutler_unified_2021}. These mocks are those used to calculate covariance matrices and contain survey geometry and systematics to match their associated measurements. Each of the galaxy surveys considered for this work has its own set of mocks. The number of mock realisations and specifics of simulations used to produce them are covered in Section 5 of \citet{beutler_unified_2021}, or for the 6dFGS mocks see \citet{koda_fast_2016,carter_low_2018}, for BOSS see \citet{klypin_multidark_2016,kitaura_clustering_2016}, and for eBOSS see \citet{chuang_ezmocks_2015,zhao_completed_2021}. It is helpful to have sets of mock multipoles for which we know the true cosmology as well as the "true" values for the nuisance parameters of the EFTofLSS model (bias parameters and counterterms, see Section \ref{sec:model}). To that end, we produce a set of mock multipoles using \textsc{PyBird} with the cosmology set to the TT,TE,EE+lowE+lensing+BAO $\Lambda$CDM best-fit values from Table 2 in \citet[][henceforth Planck 2018]{planck_collaboration_planck_2020}. The nuisance parameters are fit to the mean of the mock multipole measurements published in \citet{beutler_unified_2021} for each sample. We refer to the resulting multipoles as the "\textsc{PyBird} mocks".\\

\begin{figure}
	\includegraphics[width=\columnwidth]{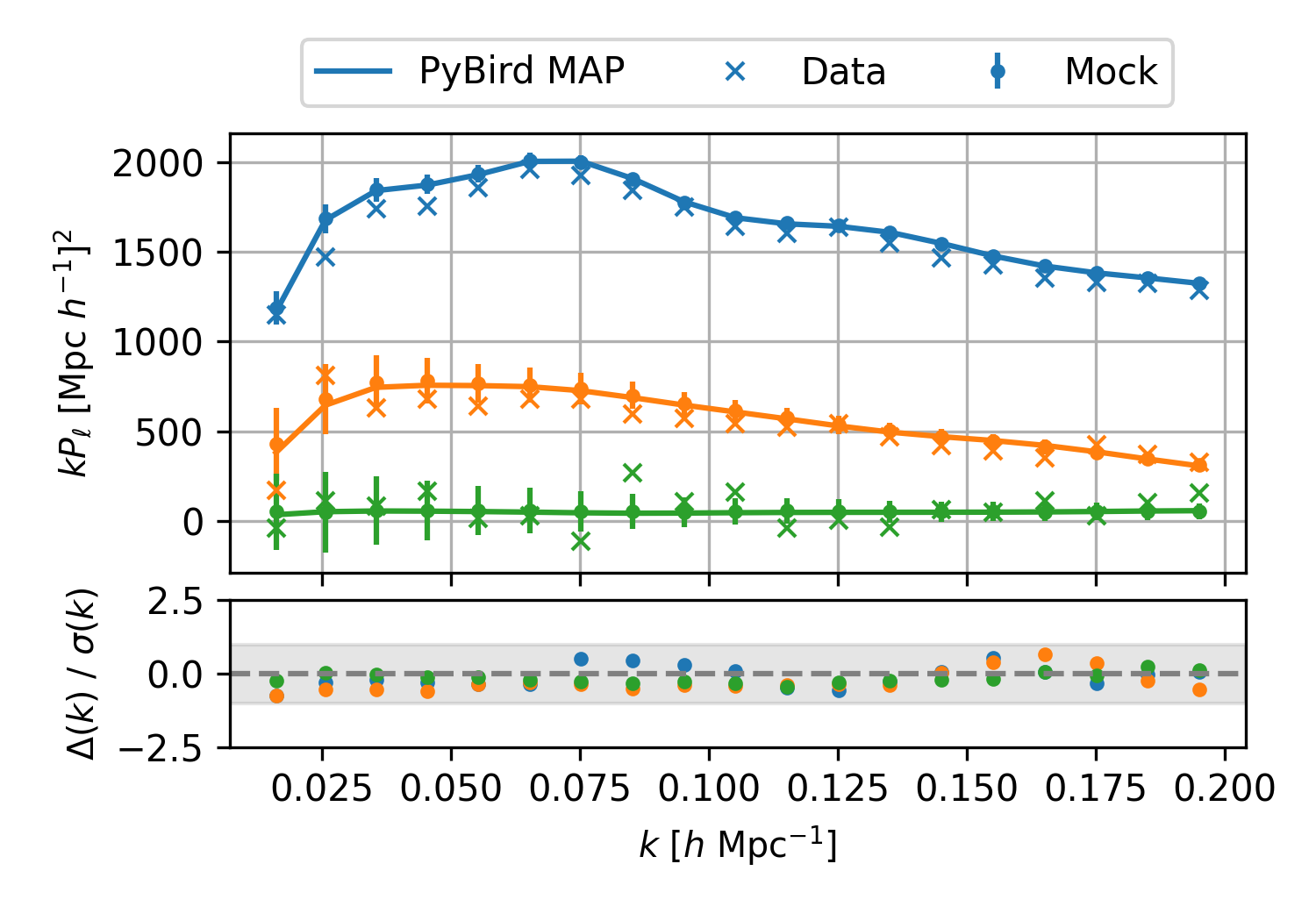}
    \caption{\textit{Top}: With points and error bars, the mean of 1049 multipoles measured from the MD-Patchy mocks \citep{kitaura_clustering_2016} for the NGC at $z=0.61$. The error bars show the $1\sigma$ error calculated from the 1049 measurements. The solid lines show the \textsc{PyBird} prediction for the Planck 2018 TT,TE,EE+lowE+lensing+BAO $\Lambda$CDM best-fit cosmology and the MAP estimate resulting from fitting bias parameters and counterterms to the mean multipoles from the MD-Patchy mocks. The crosses show the multipoles measured from BOSS NGC data, again with $z=0.61$. \textit{Bottom}: The residual of the mean multipole measurements and the \textsc{PyBird} prediction normalised by the $1\sigma$ errors reduced by a factor of 10. The colours blue, orange, and green in both panels represent the monopole, quadrupole, and hexadecapole multipole moments, respectively.}
    \label{fig:pybird_mock}
\end{figure}

The nuisance parameters for the \textsc{PyBird} mocks are determined by finding the maximum \textit{a posteriori} (MAP) estimate for four bias parameters and six counterterms. This is done by finding the minimum of the negative log-likelihood (see Section \ref{subsubsec:fid_results_mock} for likelihood definition) with a wide uniform prior on all bias parameters and counterterms. Except for the linear bias, this prior ranges from $-50 < b_i < 50$. The linear bias prior is truncated at zero to allow for positive values only. The nuisance parameters are fit to the mean of the mock multipoles on scales $0 < k < 0.2\ h\ \mathrm{Mpc}^{-1}$, and the covariance is rescaled by a factor of 10.\footnote{We rescale the covariance so that the nuisance parameters are well constrained for each sample. We could, in principle, rescale by a large factor that depends on the number of mock realisations for each sample. However, when we are producing the \textsc{PyBird} mocks, we are not looking to answer how well \textsc{PyBird} can recover different simulation methods with such large effective volumes. We are solely trying to produce synthetic multipoles that have the same functional form as the data for which all the true parameters are known.} Figure \ref{fig:pybird_mock} shows the \textsc{PyBird} mock multipoles alongside the multipole measurements and mocks from \citep{beutler_unified_2021} for the $z=0.61$ NGC sample. The bottom panel shows the residuals normalised by the rescaled covariance $\frac{\Delta(k)}{(\sigma(k)/10)}$. We can see that the agreement of the \textsc{PyBird} mock multipoles and the mocks of \citep{beutler_unified_2021} is within $1\sigma$. It should be noted that the agreement is better still when considering the unscaled covariance. Plots showing the \textsc{PyBird} mocks for the other samples all exhibit similar results.

\section{Model}
\label{sec:model}
As alluded to in Section \ref{sec:intro}, there are two general routes to modelling the galaxy power spectrum. The first is to use numerical simulations; providing accurate small-scale predictions but coming at a high computational cost. The second is to develop an analytic model; producing computationally efficient predictions (in comparison to numerical simulations) but being less accurate on small scales.\\

Probing the small, nonlinear, scales of the galaxy power spectrum can improve the constraints on the cosmological parameters. For a given survey, we will have a larger number of galaxy-galaxy pairs with small separations than large separations; thus, the statistical error on small scales will be lower than on large scales. The EFTofLSS was developed to extend the scales of validity of analytic predictions, allowing us to probe smaller scales and exploit the reduced statistical error. 

\subsection{EFTofLSS}

Standard perturbation theory (SPT) models the dark matter overdensity field as a perfect fluid. Although successful on large scales, where the density perturbations are small, its description starts to break down when entering nonlinear scales \citep[1-loop SPT breaks down at $k \sim 0.1\ h\ \mathrm{Mpc}^{-1}$ for redshift $z=0$,][]{carlson_critical_2009}. In recent years considerable effort has been put into an effective description which extends the range of SPT into a mildly nonlinear regime.\\

EFTofLSS introduces a cut-off scale which acts as an effective low-pass filter, leading to the fluid equations now being solved in terms of long-wavelength overdensity and velocity fields. Furthermore, an effective stress-energy tensor is introduced, which captures the effects of the small scales physics on the larger scales. At a given order $n$,  the effect of these small scales and their backreaction onto the long wavelength field can be captured by a finite number of so-called "counterterms" $c_i$. These counterterms are free parameters that must be fitted to data or calibrated with simulations. Including a nonlinear bias scheme, mapping the underlying dark matter field as described above to the observed galaxy densities, the 2D redshift-space galaxy power spectrum in terms of scale $k$ and cosine of angle to the line-of-sight $\mu$, can be written as
\begin{align}
     P_g (k, \mu) &= Z_1 (\mu)^2 P_{11}(k) \nonumber \\
     &+ 2 \int \frac{d^3 q}{(2 \pi)^3} Z_2(\textbf{q},\textbf{k - q}, \mu)^2 P_{11}(|\textbf{k - q}|) P_{11}(q) \nonumber \\
     &+ 6 Z_1(\mu) P_{11}(k) \int \frac{d^3 q}{(2 \pi)^3} Z_3( \textbf{q},\textbf{-q},\textbf{k },\mu) P_{11}(q) \nonumber \\
     &+ 2 Z_1(\mu) P_{11}(k) \left( c_{ct} \frac{k^2}{k_M^2} + c_{r,1} \mu^2 \frac{k^2}{k_M^2} + c_{r,2} \mu^4 \frac{k^2}{k_M^2} \right) \nonumber \\
     &+ \frac{1}{\bar{n}_g} \left( c_{\epsilon, 1} + c_{\text{mono.}} \frac{k^2}{k^2_M} + \frac{3}{2} c_{\text{quad.}} \left( \mu^2 - \frac{1}{3}  \right)   \frac{k^2}{k^2_M}\right)\ .
\end{align}
In the above $Z_i$ are the redshift-space galaxy density kernels (for their exact form, see \citealt{damico_cosmological_2020}), $\bar{n}_g$ is the mean galaxy density\footnote{For the analyses of this work we use values of $4\times10^{-4}\ h^3\ \mathrm{Mpc^{-3}}$ for the 6dFGS and BOSS samples. For the eBOSS QSO samples we use $1.5\times10^{-5}\ h^3\ \mathrm{Mpc^{-3}}$.}, and $k_M^{-1}$ is a normalisation scale\footnote{More recent papers that use the \textsc{PyBird} EFTofLSS model have an additional normalisation scale $k_R$. For this work, we neglect $k_R$, as such $k_R=k_M$. Throughout we set $k_M=0.7\ \mathrm{Mpc}^{-1}$.}. Overall the 1-loop EFTofLSS introduces ten nuisance parameters. Four parameters ($b_{1-4}$) are introduced in the expansion of the galaxy density and velocity field in terms of the underlying dark matter field. These parameters are found in the galaxy kernels $Z_i$. It has been noted that $b_2$ and $b_4$ are highly degenerate \citep{damico_cosmological_2020}. It is common to reparameterise such that
\begin{align}
    c_2 = (b_2+b_4)\ / \ \sqrt{2}\ , \nonumber \\
    c_4 = (b_2-b_4)\ / \ \sqrt{2}\ .
\end{align}
There are three stochastic parameters ($c_{\epsilon, 1},c_{\text{mono.}},c_{\text{quad.}}$) that are introduced to capture the difference between the actual observed galaxy field and its expected value. Finally, three counterterms that encapsulate the impact of UV physics: the effective sound speed of the dark matter field $c_{ct}$, and $c_{r,1}$ and $c_{r,2}$ which control the impact of small scales on redshift space distortion.

\subsection{Alcock-Paczyński effect}

A reference cosmology is required to measure the galaxy power spectrum from redshift catalogues provided by surveys like those introduced in Section \ref{sec:data}. Any differences between the true underlying cosmology and the reference cosmology lead to distortions of distances parallel and perpendicular to the line of sight. This is the so-called Alcock-Paczyński (AP) effect \citep{alcock_evolution_1979}. The distortion parallel and perpendicular to the line of sight is given by the distortion parameters $q_\parallel$ and $q_\bot$, respectively. These parameters are defined as
\begin{align}
    q_\parallel = \frac{D_A(z)H(z=0)}{D_A^\mathrm{ref.}(z)H(z=0)}\ , \nonumber \\
    q_\bot = \frac{H^\mathrm{ref.}(z)H(z=0)}{H(z)H^\mathrm{ref.}(z=0)}\ ,
\end{align}
with $H(z)$ and $D_A(z)$ being the Hubble parameter and angular-diameter distance as a function of redshift, respectively. The superscript ref. in the above equations indicates quantities calculated at the reference cosmology. The AP distortion is applied to the scales and angles as $k'=q_\bot^{-1} B k^\mathrm{ref.}$ and $\mu'=F^{-1} B^{-1} \mu^\mathrm{ref.}$. With $F=q_\parallel\ / \ q_\bot$, and $B$ given by
\begin{equation}
    B = \left[1+\left(\mu^\mathrm{ref.}\right)^2\left(F^{-2}-1\right)\right]^{1/2}\ .
\end{equation}
The 2D power spectrum can then be decomposed into multipoles via
\begin{equation}
    P_l(k) = \frac{2l+1}{2q_\parallel q_\bot^2}\int_{-1}^1 P\left(k', \mu'\right)\mathcal{L}_l\left(\mu^\mathrm{ref.}\right)\mathrm{d}\mu^\mathrm{ref}\ ,
    \label{eq:AP_multi}
\end{equation}
with $\mathcal{L}_l$ being the $l$-th order Legendre polynomial.\\

The \textsc{EFTEMU} (and \textsc{PyBird}) make predictions for the power spectrum multipoles rather than the 2D power spectrum. To include the AP effect, via Equation \ref{eq:AP_multi}, we need to reconstruct the 2D power spectrum from the multipoles. We do this via
\begin{equation}
    P(k, \mu) = \sum_{l=0} P_l(k)\mathcal{L}_l(\mu)\ .
    \label{eq:recon_2D}
\end{equation}
The \textsc{EFTEMU} (as trained for this work) makes predictions for the first two even multipoles. Reconstructing the 2D power spectrum from only the first two even multipoles will result in systematic errors when including the AP effect via Equation \ref{eq:AP_multi}. These errors are expected to be small compared to the error associated to the multipole measurements discussed in Section \ref{sec:data}. It should be noted that the \textsc{PyBird} mocks introduced in Section \ref{subsubsec:pybird_mocks} were constructed including the hexadecapole $P_4(k)$. As such, the mock analyses of Section \ref{sec:mock_analyses} will test if these systematic errors from the 2D power spectrum reconstruction impact the inferred cosmology.
\\

\subsection{Emulator}
\label{subsec:eftemu}

The EFTofLSS model described above (as implemented in \textsc{PyBird}) takes $\mathcal{O}$(1 s) to produce predictions for a given set of cosmological parameters at a given redshift. Although efficient enough for direct use when conducting cosmological inference, this prediction time does prohibit the exploration of analysis setups (such as prior choice, scale cuts, and fixed parameters). If running a typical MCMC using this model requires $\mathcal{O}(10^5 \text{--} 10^6)$ model evaluations, then $\mathcal{O}$(days) would be required to reach convergence. In \citet{donald-mccann_matryoshka_2022}, the \textsc{EFTEMU} was added to the \textsc{matryoshka} \citep{donald-mccann_span_2022} suite of emulators. The \textsc{EFTEMU} was developed to accelerate EFTofLSS predictions by several orders of magnitude by replacing the direct calculation of the kernels $P_{n,l}$ of the EFTofLSS model with predictions from simple NNs.\\

\begin{figure*}
    \centering
    \includegraphics[width=\linewidth]{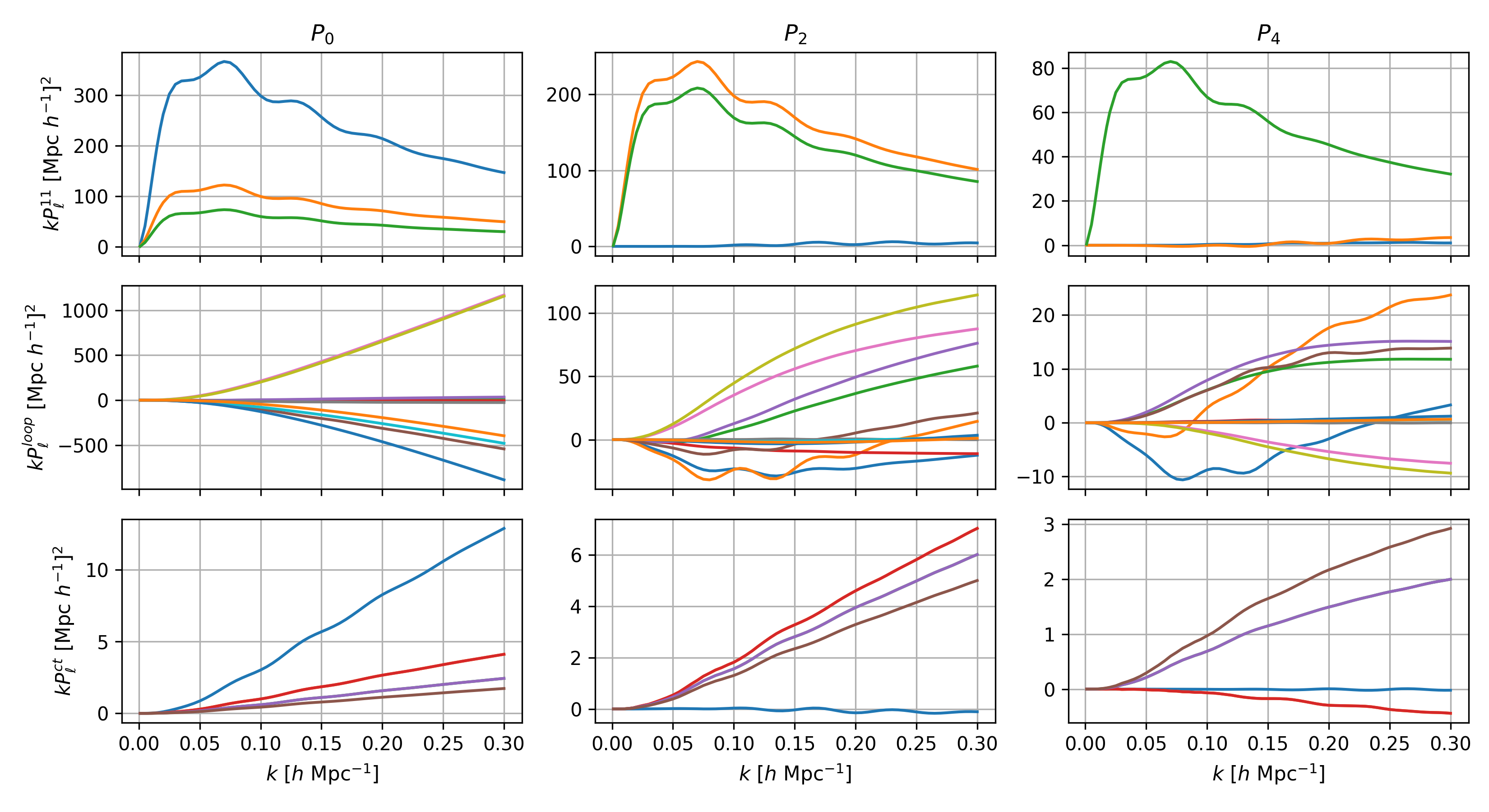}
    \caption{Redshift space kernels $P_{n,l}$ calculated with \textsc{PyBird} for the Planck 2018 TT,TE,EE+lowE+lensing+BAO $\Lambda$CDM best-fit cosmology at $z=0.61$.}
    \label{fig:pybird_mock_kernels}
\end{figure*}

The \textsc{EFTEMU} was originally trained with data drawn from a five-dimensional $\Lambda$CDM parameter space, approximately centred on the Planck 2018 best-fit cosmology. Despite being wide, this training space is too restrictive to constrain some of the $\Lambda$CDM parameters much beyond this when using the large-scale structure data considered for this work. With this in mind, we re-train the \textsc{EFTEMU} for this work. The width of the prior on $\omega_c$, $h$, and $\ln{(10^{10}A_s)}$ was increased significantly, and the spectral index $n_s$ was fixed as we do not expect to get any meaningful constraint on $n_s$ from our analyses. Table \ref{tab:emu_priors} compares the prior for the original \textsc{EFTEMU} to that used in this work. The larger training space required a change in the training procedure compared to that in \citet{donald-mccann_matryoshka_2022}. The increased width of the cosmological prior, particularly for $\ln{(10^{10}A_s)}$, increases the dynamic range of the kernels $P_{n,l}$. The original preprocessing procedure involved rescaling all $P_{n,l}$ such that at every $k$-value their magnitude was in the range $[0,1]$. We modify this procedure by first taking the log of the $P_{n,l}$ before rescaling into the range $[0,1]$. Figure \ref{fig:pybird_mock_kernels} shows the kernels for the \textsc{PyBird} mocks at $z=0.61$ for the first three even multipoles on scales $0.001 \leq k \leq 0.3 \ h \ \mathrm{Mpc}^{-1}$. There are 21 kernels for each multipole, and these 21 kernels can be split into three groups. The first group ($P_{n,l}^\mathrm{11}$) contains the linear terms, the second group ($P_{n,l}^\mathrm{loop}$) contains the loop terms, and the third group ($P_{n,l}^\mathrm{ct.}$) contains the counterterms. These three groups also represent the grouping used for the \textsc{EFTEMU}; each component of the \textsc{EFTEMU} emulates a different group \citep[see Section 3 of][]{donald-mccann_matryoshka_2022}. It can be seen from Figure \ref{fig:pybird_mock_kernels} that some of the $P^\mathrm{loop}_l$ and $P^\mathrm{ct.}_l$ kernels are exclusively negative or have a zero crossing. To allow us to take the log of these kernels, we include either a simple sign change or the addition of a constant to the kernel preprocessing. Taking the log results in a reduced dynamic range in the training data and leads to higher prediction accuracy. We also significantly increase the number of samples generated for training and testing from 10,000 to 50,000. Only 40,000 are used for training; the remaining 10,000 are used for testing.\\

\begin{table}
 \centering
 \begin{tabular}{c c c} 
 \hline
 Parameter & \citet{donald-mccann_matryoshka_2022} & This Work\\ 
 \hline
 $\omega_c$ & $\mathcal{U}(0.101, 0.140)$ & $\mathcal{U}(0.0900, 0.160)$ \\
 $\omega_b$ & $\mathcal{U}(0.0210, 0.0240)$ & $\mathcal{U}(0.0200, 0.0240)$ \\
 $h$ & $\mathcal{U}(0.575, 0.748)$ &  $\mathcal{U}(0.500, 0.850)$ \\
 $\ln{\left(10^{10}A_s\right)}$ & $\mathcal{U}(2.78, 3.32)$ & $\mathcal{U}(1.50, 3.75)$\\
 $n_s$ & $\mathcal{U}(0.901, 1.03)$ & 0.965 \\
 \hline
 \end{tabular}
\caption{Comparison of priors on the cosmological parameters of the \textsc{EFTEMU} from \citet{donald-mccann_matryoshka_2022} and this work. $\mathcal{U}(a, b)$ denotes a uniform distribution with boundaries $a$ and $b$.}
\label{tab:emu_priors}
\end{table}

\begin{figure}
	\includegraphics[width=\linewidth]{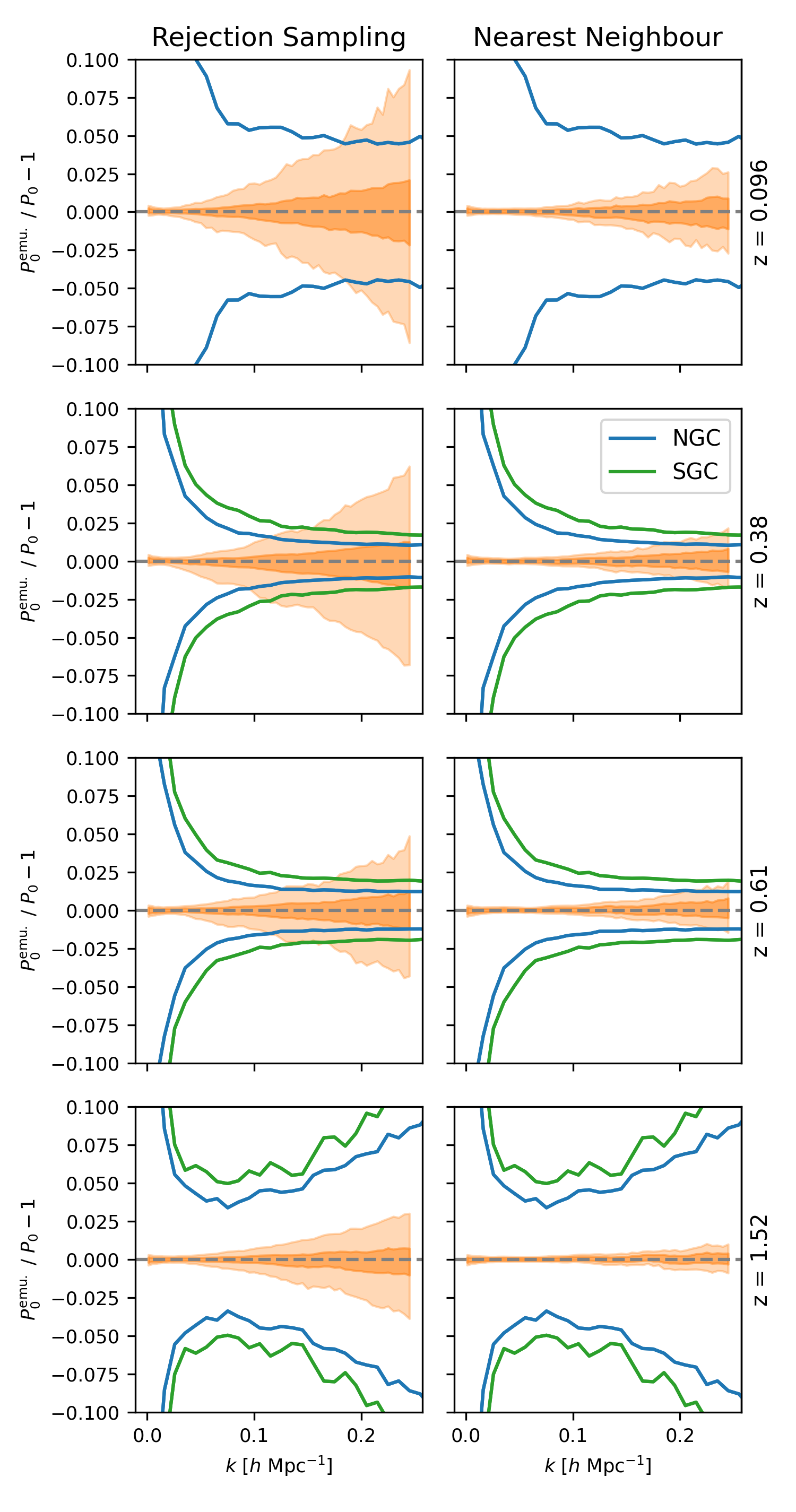}
    \caption{Prediction error of the re-trained \textsc{EFTEMU} used in this work. The orange shaded regions in each panel show the 68\% and 95\% credible intervals of the prediction error, respectively. The credible intervals are calculated by examining the prediction error on 10,000 test cosmologies not used for training. The prediction error is defined as the ratio of the \textsc{EFTEMU} prediction to the \textsc{PyBird} prediction for the same set of cosmological and nuisance parameters. The ratio is then normalised such that it is equal to zero for a perfect prediction. Each row represents a different redshift 0.096, 0.38, 0.61, and 1.52 from top to bottom. For the left column, the cosmological parameters are combined with random draws of nuisance parameters from the theoretically viable prior space. For the right column, each test cosmology is combined with a set of nuisance parameters that result in 6dFGS-like predictions. The coloured solid lines show the inverse signal-to-noise ratio on the monopole for the datasets considered for this work. Panels with both blue and green lines represent the NGC and SGC, respectively.}
    \label{fig:per_err}
\end{figure}

Figure \ref{fig:per_err} shows the prediction error on the monopole of the power spectrum when producing predictions with the re-trained \textsc{EFTEMU}. Each row shows the prediction error at a different redshift, and each column shows the prediction error computed with different sets of nuisance parameters. The orange shaded regions show the 68\% and 95\% credible intervals (CIs) of the prediction error as a function of $k$. The solid coloured lines show the inverse signal-to-noise ratio (SNR) for the monopole measurements considered for this work at their respective redshifts. The shaded regions have been calculated from predictions for 10,000 unseen cosmologies. For the left column, the 10,000 cosmologies have been combined with sets of nuisance parameters that produce "reasonable" predictions for the monopole. We take random draws from a very wide uniform prior\footnote{$0<b_1<10$, $-10<\left\{b_2,\ b_4\right\}<10$, $-500<\left\{b_3,\ c_{ct},\ c_{r,1},\ c_{r,2}\right\}<500$.} on the nuisance parameters and calculate the multipoles for each set of cosmological and nuisance parameters. We define "reasonable" predictions as those which the monopole is strictly positive and those which can be said to remain perturbative\footnote{See Appendix \ref{app:perturb_cond} for our perturbative condition.}. Any sets of parameters that do not meet these criteria are rejected, and the nuisance parameters resampled from the prior. This is repeated until we have nuisance parameters for all 10,000 cosmologies. For the right column, samples from the posterior resulting from full-shape analysis of the 6dFGS-like \textsc{PyBird} mock (see Section \ref{sec:mock_analyses}) are used to inform the nuisance parameters for the unseen cosmologies. For each unseen test cosmology, the posterior sample with the closest cosmology\footnote{The nearest neighbour in the 4D cosmological parameter space. With the Euclidean distance as the distance metric.} is selected, and its nuisance parameters are associated to that test cosmology. The two columns of Figure \ref{fig:per_err} show two different aspects of the prediction accuracy: the left column represents the prediction accuracy across the entire theoretically viable parameter space, the right column represents the prediction accuracy for power spectra that look more similar to something that has been previously observed. We can see from the right column that for all redshifts considered and for all $k<0.25\ h\ \mathrm{Mpc}^{-1}$, the prediction error from the emulator is less than the error on the data at the 68\% level at each respective redshift. However, from the left column, we can see that for $z=0.38,\ 0.61$ when considering the entire theoretically viable prior space, the prediction error can be greater than the error on the data on small scales ($k\gtrsim 0.17\ h\ \mathrm{Mpc}^{-1}$). In practice, we find that the level of prediction accuracy from the re-trained \textsc{EFTEMU} does not induce any significant bias to the cosmological parameters when performing inference, as shown in Section \ref{sec:mock_analyses}.

\section{Mock Analyses}
\label{sec:mock_analyses}
In this section, we present the results from a series of analyses of the \textsc{PyBird} mocks (described in Section \ref{subsubsec:pybird_mocks}). These mock analyses aim to verify that our cosmological inference pipeline does not induce biases in the cosmological parameter constraints. In addition, we explore how various analysis setups impact the results. In all cases, to put constraints on cosmological parameters, we sample from the posterior distribution via \textit{Preconditioned Monte Carlo} \citep{karamanis_accelerating_2022}; as implemented in \textsc{pocoMC}\footnote{Various parameters control the efficiency of the sampling with \textsc{pocoMC}. We use the default values for all of these.} \citep{karamanis_pocomc_2022}. Precondition Monte Carlo utilises \textit{Normalising Flows} \citep{papamakarios_normalizing_2021} and \textit{Sequential Monte Carlo} \citep{del_moral_sequential_2006} to efficiently sample from posterior distributions even when they have a very complex shape. We use a Gaussian likelihood of the form
\begin{equation}
    \ln{\left[\mathcal{L}(P|\theta,\phi)\right]} = -\frac{1}{2}(P-\tilde{P})^T\bm{C}^{-1}(P-\tilde{P})\ ,
    \label{eq:likelihood}
\end{equation}
with $P$ being a concatenation of the multipole measurements considered $P=[P_0, P_2]$, $\tilde{P}$ being the multipole predictions from the model $\tilde{P}=[\tilde{P}_0, \tilde{P}_2]$ for a given set of cosmological parameters $\theta$ and nuisance parameters $\phi$, and $\bm{C}$ being the covariance matrix.\\

Many of the nuisance parameters of the EFTofLSS model appear linearly as multiplicative factors for the kernels. This allows us to marginalise over these parameters analytically rather than sampling from them. This is standard practice when conducting parameter inference with the EFTofLSS \citep{damico_cosmological_2020, damico_limits_2021,glanville_full-shape_2022}. Carrying out the analytic marginalisation reduces dimensionality and thus leads to a more efficient inference of the cosmological parameters. Although it is more efficient to analytically marginalise the linearly appearing parameters, the prediction speed of the \textsc{EFTEMU} means that fully sampling the parameter space is tractable. We refer to the likelihood with no analytic marginalisation as the "full" likelihood, and we explore the use of both the marginalised and full likelihood in the results below.

\subsection{Fiducial Results}
\label{subsubsec:fid_results_mock}

We start by presenting results from an analysis with a fiducial setup. For this fiducial setup, we analyse the power spectrum monopole and quadrupole on scales $0.01 < k < 0.15 \ h\ \mathrm{Mpc}^{-1}$. Figure \ref{fig:per_err} shows that the nearest neighbour prediction error on these scales is considerably lower than the error associated to the mocks at all redshifts for which the \textsc{EFTEMU} is trained. We fix three out of the ten nuisance parameters to zero, those parameters being $c_4$, $c_{r,2}$, $c_\mathrm{mono.}$. These parameters are commonly set to zero in analyses of the monopole and quadrupole with \textsc{PyBird} \citep{damico_cosmological_2020,simon_consistency_2022}. The priors on $\omega_c$, $h$, and $\ln{\left(10^{10}A_s\right)}$ are those that define the emulator training space (given in Table \ref{tab:emu_priors}). For $\omega_b$, we use a truncated normal distribution as the prior, with a mean of 0.02235 and a standard deviation of 0.00049\footnote{This is motivated by BBN \citep{cooke_one_2018} and are the same values as those used in \citet{glanville_full-shape_2022}.}. The hard bounds of this prior are given by the emulator training space as with the other cosmological parameters. The priors on the nuisance parameters are given in Table \ref{tab:bias_priors}. We refer to the prior of Table \ref{tab:bias_priors} as the "classic" prior. A majority of the EFTofLSS works cited in this paper use a prior of a similar form. Note that the prior on $c_{\epsilon,1}$ is defined independent of $\bar{n}_g$. For $\bar{n}_g=4\times10^{-4} h^3\ \mathrm{Mpc}^{-3}$ the prior width is 400, which is in line with other works that use the \textsc{PyBird} EFTofLSS model.\\

\begin{figure*}
	\includegraphics[width=\linewidth]{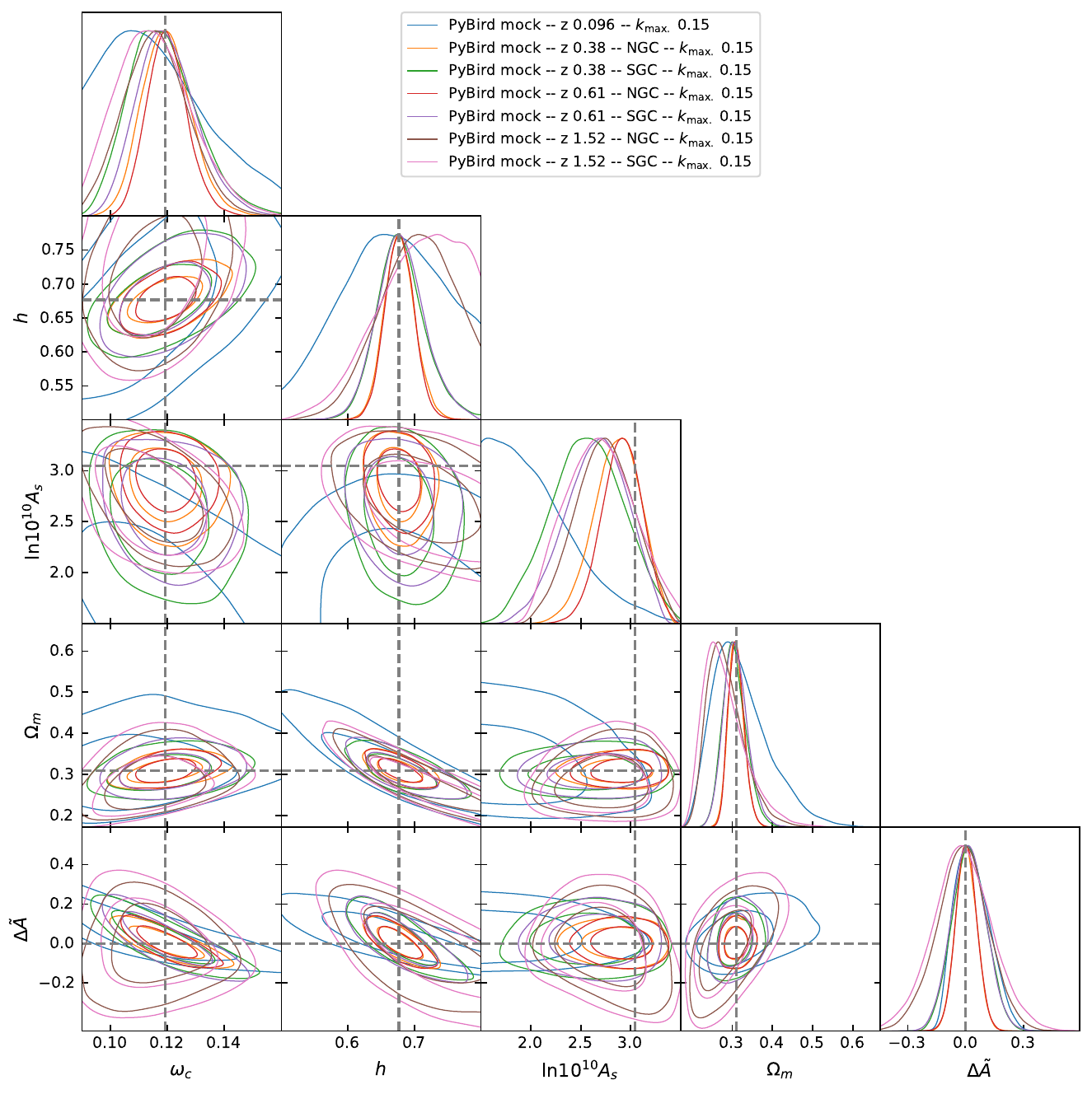}
    \caption{1D and 2D marginalised posterior distributions on the cosmological parameters of interest resulting from analysis of the \textsc{PyBird} mocks with the fiducial analysis setup (described in Section \ref{subsubsec:fid_results_mock}). The two contour levels in the off-diagonal panels represent the $1\sigma$ and $2\sigma$ regions, and the grey dashed lines in all panels show the true values of the \textsc{PyBird} mocks. The parameters $\Omega_m$ and $\tilde{A}$ have been derived, whilst the other parameters were sampled (see Section \ref{subsubsec:fid_results_mock} for details).}
    \label{fig:corner_pybird_mock_model1}
\end{figure*}

Figure \ref{fig:corner_pybird_mock_model1} shows the resulting marginalised 1D and 2D posteriors from the analysis of the \textsc{PyBird} mocks with the fiducial setup and using the full likelihood\footnote{Throughout this work, plots showing marginalised posterior distributions have been produced directly or with the assistance of \textsc{GetDist} \citep{lewis_getdist_2019}.}. The two contour levels in the off-diagonal panels are $1\sigma$ and $2\sigma$, and the grey dashed lines indicate the location of the true values used to generate the mocks. Along with the sampled parameters $\omega_c$, $h$, and $\ln{\left(10^{10}A_s\right)}$ we also plot the marginalised posterior distributions on two derived parameters: $\Omega_m=(\omega_c+\omega_b)h^{-2}$, and $\tilde{A}=b_1^2A_s10^8$. For the purposes of this plot, the derived $\tilde{A}$ posterior samples have had the truth subtracted, such that the 1D marginalised posterior should peak exactly at zero if unbiased. This normalisation of $\tilde{A}$ allows us to compare the distributions calculated for each sample as they all have different $b_1$ values. Looking at Figure \ref{fig:corner_pybird_mock_model1}, it is clear that for \textsc{PyBird} mocks with a higher SNR (BOSSz1 and BOSSz3 NGC), the agreement with the truth is very good for all parameters. For \textsc{PyBird} mocks with a lower SNR (6dFGS and eBOSS QSO SGC), we observe some significant shifts from the truth in many of the 1D and 2D projections. A likely cause for these shifts is the \textit{volume effect} \citep{carrilho_cosmology_2022,simon_consistency_2022,hadzhiyska_cosmology_2023}; these shifts are (at least partially) a result of marginalisation. In previous works, it has been shown that $\ln{\left(10^{10}A_s\right)}$ is particularly susceptible to volume effects \citep{carrilho_cosmology_2022, simon_consistency_2022}, and indeed it is the parameter in Figure \ref{fig:corner_pybird_mock_model1} that shows the most significant observed shift. See Appendix \ref{app:toy} for more discussion on the volume effect with a toy example.\\ 

\begin{table}
 \centering
 \begin{tabular}{c c c c} 
 \hline
 Parameter & Prior & $\mathcal{M}_1$ & $\mathcal{M}_3$\\ 
 \hline
 $b_1$ & $\mathcal{U}(0, 4)$ & \checkmark & \checkmark\\
 $c_2$ & $\mathcal{U}(-4, 4)$ & \checkmark & \checkmark\\
 $b_3$ & $\mathcal{N}(0, 2)$ & \checkmark& \scalebox{0.75}{\usym{2613}}\\
 $c_4$ & $\mathcal{N}(0, 2)$ & \scalebox{0.75}{\usym{2613}} & \scalebox{0.75}{\usym{2613}}\\
 $c_{ct}$ & $\mathcal{N}(0, 2)$ & \checkmark& \scalebox{0.75}{\usym{2613}}\\
 $c_{r,1}$ & $\mathcal{N}(0, 8)$ & \checkmark & \checkmark\\
 $c_{r,2}$ & $\mathcal{N}(0, 2)$ & \scalebox{0.75}{\usym{2613}} & \scalebox{0.75}{\usym{2613}}\\
 $c_{\epsilon,1}$ & $\mathcal{N}(0, 0.16)$ & \checkmark& \scalebox{0.75}{\usym{2613}}\\
 $c_\mathrm{mono.}$ & $\mathcal{N}(0, 2)$ & \scalebox{0.75}{\usym{2613}} & \scalebox{0.75}{\usym{2613}}\\
 $c_\mathrm{quad.}$ & $\mathcal{N}(0, 2)$ & \checkmark& \scalebox{0.75}{\usym{2613}}\\
 \hline
 \end{tabular}
\caption{Priors on the bias parameters and counterterms of the \textsc{PyBird} EFTofLSS model. $\mathcal{U}(a, b)$ denotes a uniform distribution with boundaries $a$ and $b$, and $\mathcal{N}(\mu, \sigma)$ denotes a normal distribution with mean $\mu$ and standard deviation $\sigma$. The last two columns indicate which parameters are included in the two sub-models $\mathcal{M}_1$ and $\mathcal{M}_3$ (defined in Section \ref{subsubsec:model_comp}).}
\label{tab:bias_priors}
\end{table}

\begin{figure}
	\includegraphics[width=0.95\linewidth]{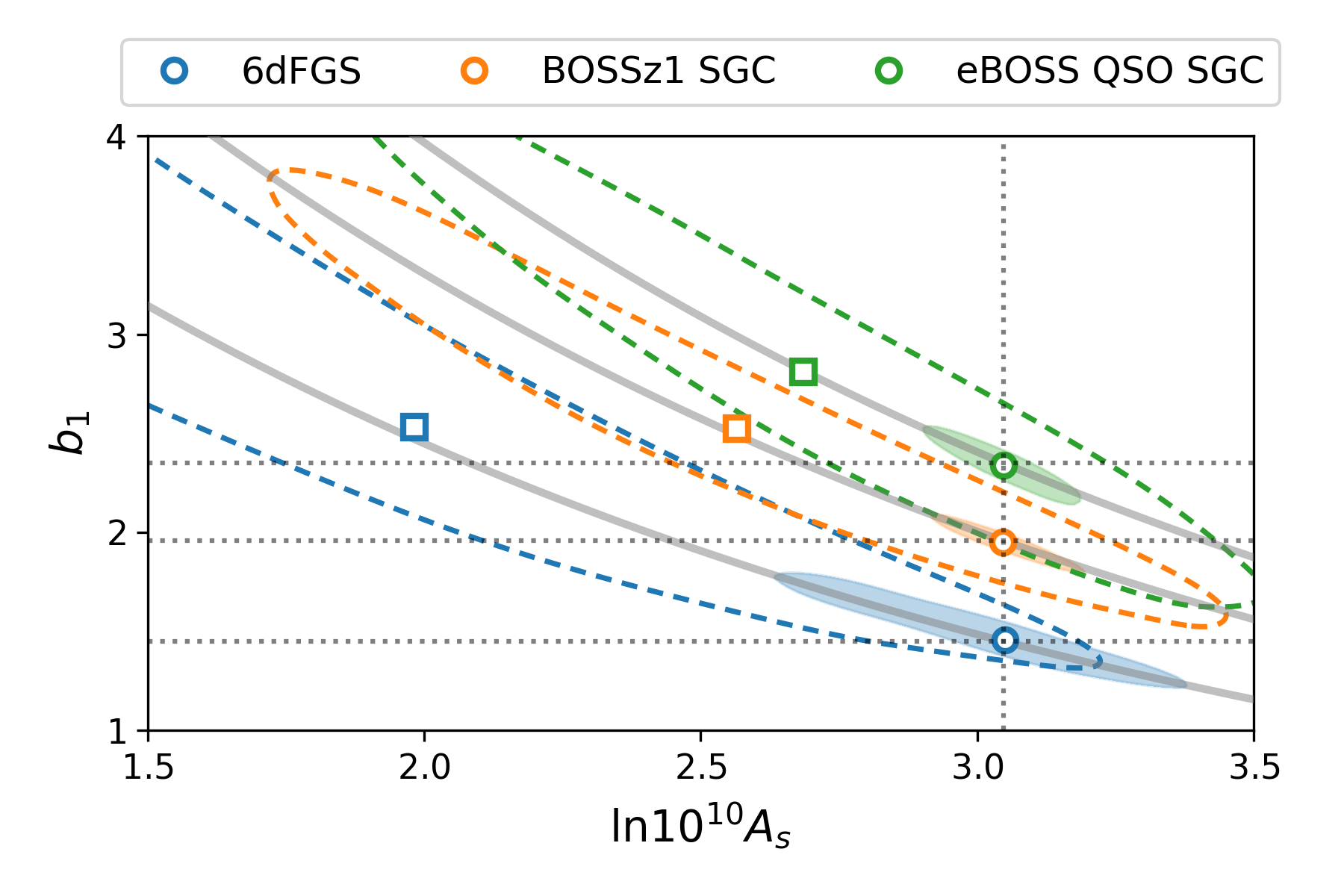}
    \caption{2D marginalised posterior for $b_1$ and $\ln{\left(10^{10}A_s\right)}$ resulting from analysis of mocks representing various samples of interest for this work with the fiducial setup described in Section \ref{subsubsec:fid_results_mock}. The dashed coloured contours represent the $2\sigma$ region calculated when analysing mocks with covariance representative of their respective datasets. The filled contours represent the $2\sigma$ region calculated with the covariance rescaled by a factor of 50. The coloured squares show the median values of the posterior obtained from analysis with the standard covariance and the circles from analysis with the rescaled covariance. The vertical dotted line shows the true $\ln{\left(10^{10}A_s\right)}$ value of the mock, and the horizontal dashed lines show the true $b_1$ values for each mock. The grey solid lines show lines of constant $\tilde{A}$ with $b_1$ values equal to the truth from the mocks.}
    \label{fig:b1_As_norm_cov}
\end{figure}

The shifts induced in marginalised posteriors are reduced when the constraining power from the data is higher. Figure \ref{fig:b1_As_norm_cov} shows, with dashed coloured lines, the $2\sigma$ region of the 2D marginalised posterior distributions on $b_1$ and $\ln{\left(10^{10}A_s\right)}$ resulting from analysis of the \textsc{PyBird} mocks for various samples with the fiducial setup described above. Also plotted in Figure \ref{fig:b1_As_norm_cov}, with coloured
shaded regions, is the $2\sigma$ region of the 2D marginalised posteriors obtained from analysis of the \textsc{PyBird} mocks with covariance matrices rescaled by a factor of $1 \ / \ 50$. It can be seen that although there is agreement with the truth (represented with dotted grey lines) at the $2\sigma$ level in both cases for all the data samples plotted, the agreement is significantly better when the covariance has been rescaled. The posteriors have shrunk and remained consistent with the truth. If it were the case that the biases observed in Figure \ref{fig:corner_pybird_mock_model1} were resulting from anything other than marginalisation, we would not see this behaviour. We also note from Figure \ref{fig:b1_As_norm_cov} that the shift in posteriors and median values (shown with coloured squares and points) resulting from rescaling the covariance is along a line of constant $\tilde{A}$ (shown with grey solid lines). Giving a compelling argument for using $\tilde{A}$ as a diagnostic quantity when understanding if observed biases in   $\ln{\left(10^{10}A_s\right)}$ are a result of a true systemic bias from the analysis pipeline or a result of volume effects. Finally, we note that rescaling the covariance in this way does not only resolve the observed bias in $\ln{\left(10^{10}A_s\right)}$, but in all parameters shown in Figure \ref{fig:corner_pybird_mock_model1}. 

\subsection{Exploration of Analysis Setups}

The results from the previous section have shown that the analysis pipeline developed for this work can return unbiased constraints on cosmological parameters of interest for a typical EFTofLSS analysis setup. We can exploit the increased prediction speed of the \textsc{EFTEMU} to explore various analysis setups and observe their impact on the constrained cosmology.

\subsubsection{Scale Cuts}
\label{subsubsec:scale_cuts}

We start by exploring different scale cuts. It can be seen from the solid coloured lines in Figure \ref{fig:per_err} that there is clear scale dependence in the inverse SNR for all the data samples considered for this work. There is also a clear scale dependence in the emulator prediction error. As mentioned in Section \ref{sec:model}, when analysing LSS data, there is a general expectation that the SNR increases when pushing to smaller scales. However, this is only true if the scales are not dominated by shot noise. If we combine this with a higher modelling error on smaller scales, although the expectation might be that including smaller scales will improve the constraints, this might not be the case.\\

\begin{figure*}
	\includegraphics[width=0.95\linewidth]{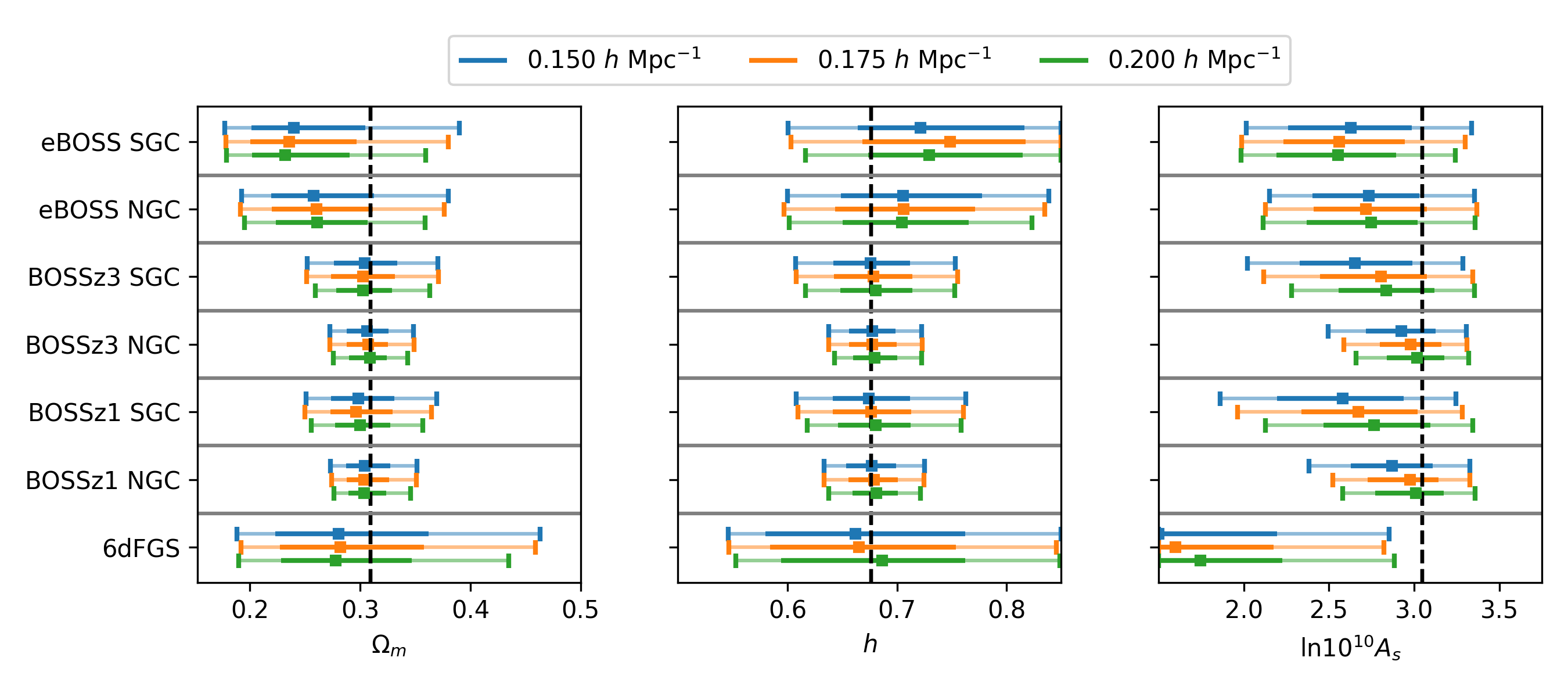}
    \caption{Summary of the 1D marginalised posteriors on cosmological parameters of interest resulting from the analyses described in Section \ref{subsubsec:scale_cuts}. Coloured squares show peak posterior values, dark horizontal coloured lines show the width of the 68\% CI, light coloured lines with caps show the 95\% CI, and vertical dashed lines show the true values of the mocks.}
    \label{fig:model1_kmax_comp}
\end{figure*}

Figure \ref{fig:model1_kmax_comp} shows the peak posterior values and 68\% CIs of 1D marginalised posteriors (with coloured squares and lines, respectively) on the cosmological parameters $\Omega_m$, $h$, and $\ln{\left(10^{10}A_s\right)}$ resulting from analysis of the \textsc{PyBird} mocks with $k_\mathrm{max.}=0.150,0.175,0.200\ h\ \mathrm{Mpc}^{-1}$ and the full likelihood. The results from the analysis of the BOSS-like mocks all show the same general trend; including smaller scales shrinks the 68\% CI, reduces the observed bias in the peak posterior value, or both. The results for 6dFGS show a slightly tighter constraint on $\Omega_m$ and $h$ when including smaller scales but the constraint on $\ln{\left(10^{10}A_s\right)}$ remains almost constant. This is likely because the constraint on $\ln{\left(10^{10}A_s\right)}$ from 6dFGS is completely dominated by volume effects. We can also see that including smaller scales worsens the agreement with the truth for the eBOSS-like mocks; the 68\% CI shrinks, the peak posterior shifts away from the truth, or both. As can be seen from Figure \ref{fig:per_err}, the emulator error is always significantly lower than the error associated with the eBOSS-like mocks; thus, the cause for the behaviour of the eBOSS-like results is more likely to be a result of the worsening SNR rather than emulator error. It can also be seen from  Figure \ref{fig:per_err} that the smaller-scale modes have larger errors, thus including them worsens the volume effect.\\

Table \ref{tab:model1_kmax_tension} quantifies the level of agreement between the true cosmological parameters of the \textsc{PyBird} mocks and the 1D marginalised posteriors resulting from analysis of these mocks with $k_\mathrm{max.}=0.15\ h\ \mathrm{Mpc}^{-1}$ and $k_\mathrm{max.}=0.2\ h\ \mathrm{Mpc}^{-1}$. For the purposes of this paper, we quantify the agreement as the number of $\sigma$ separating the peak posterior values of two given marginalised distributions. We define the agreement $N_\sigma$ as
\begin{equation}
    N_\sigma = \frac{| \mu_0 - \mu_i |}{\sqrt{\sigma_0^2+\sigma_i^2}}\ ,
    \label{eq:tension}
\end{equation}
with $\mu_i$ and $\sigma_i$ being the mean and $1\sigma$ error calculated from the 1D marginalised posterior, and $\mu_0$ and $\sigma_0$ being the mean and $1\sigma$ error of the reference (when calculating $N_\sigma$ for the \textsc{PyBird} mocks $\sigma_0=0.$). In the case of asymmetric distributions, if the residual $\mu_0 - \mu_i$ is positive, we use the $1\sigma$ error to the right of the peak posterior. If the residual is negative, we use the $1\sigma$ error to the left of the peak posterior. We note that for all apart from the eBOSS-like mocks, the level of agreement does not significantly change and is at the $\lesssim 0.5\sigma$ level for $\Omega_m$ and $h$ when comparing the results obtained with the two $k_\mathrm{max.}$ values. For the BOSS-like mocks, the level of agreement improves to $<1\sigma$ for $\ln{\left(10^{10}A_s\right)}$ when including smaller scales. It is also worth noting that although the analyses with $k_\mathrm{max.}=0.2\ h\ \mathrm{Mpc}^{-1}$ include scales at which the observed emulator error from Figure \ref{fig:per_err} is at a similar level to the data error, we find no significant bias in the constrained cosmology for those samples least susceptible to volume effects (BOSSz1 NGC, BOSSz3 NGC).

\begin{table}
    \centering
    \begin{tabular}{c c c c c c c} 
    \hline
    Sample & \multicolumn{2}{|c|}{$\Omega_m$} & \multicolumn{2}{|c|}{$h$} & \multicolumn{2}{|c|}{$\ln{\left(10^{10}A_s\right)}$}\\ 
    \hline
    6dFGS & \textbf{0.36} & 0.46 & 0.14 & \textbf{0.11} & \textbf{2.23} & 2.71 \\ 
    BOSSz1 NGC & \textbf{0.24} & 0.3 & \textbf{0.01} & 0.21 & 0.74 & \textbf{0.22} \\ 
    BOSSz1 SGC & \textbf{0.35} & 0.35 & \textbf{0.07} & 0.12 & 1.31 & \textbf{0.86} \\ 
    BOSSz3 NGC & 0.16 & \textbf{0.05} & \textbf{0.03} & 0.14 & 0.61 & \textbf{0.19} \\ 
    BOSSz3 SGC & \textbf{0.19} & 0.27 & \textbf{0.02} & 0.12 & 1.17 & \textbf{0.75} \\ 
    eBOSS NGC & \textbf{0.94} & 1.07 & \textbf{0.51} & 0.52 & \textbf{1.06} & 1.09 \\ 
    eBOSS SGC & \textbf{1.08} & 1.33 & \textbf{0.78} & 0.95 & \textbf{1.17} & 1.45 \\ 
    \hline
    \end{tabular}
\caption{Number of sigma between the true cosmology of the \textsc{PyBird} mocks and the 1D marginalised posteriors, resulting from analysis with $k_\mathrm{max.}=0.15\ h\ \mathrm{Mpc}^{-1}$ and $k_\mathrm{max.}=0.2\ h\ \mathrm{Mpc}^{-1}$. Left and right columns for each cosmological parameter correspond to $k_\mathrm{max.}=0.15\ h\ \mathrm{Mpc}^{-1}$ and $k_\mathrm{max.}=0.2\ h\ \mathrm{Mpc}^{-1}$, respectively. Lower values are indicated with bold font.}
\label{tab:model1_kmax_tension}
\end{table}

\subsubsection{Bayesian Model Comparison}
\label{subsubsec:model_comp}

\textsc{pocoMC} allows us to easily calculate the Bayesian evidence for each posterior distribution. We use these evidence calculations to compare EFTofLSS \textit{sub-models}. We define the \textit{full model} as the \textsc{PyBird} EFTofLSS model with all nuisance parameters free, and a \textit{sub-model} as any model that results from fixing any single nuisance parameter or combination of parameters to zero.\\

\begin{figure}
	\includegraphics[width=\columnwidth]{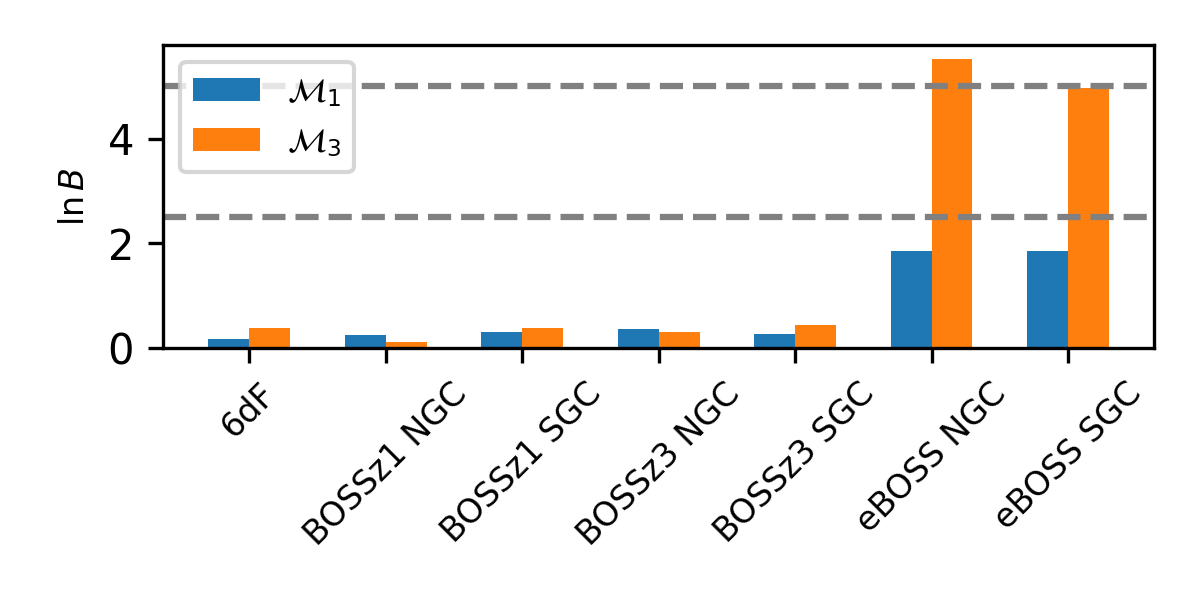}
    \caption{Natural log of the Bayes factor comparing two EFTofLSS sub-models, $\mathcal{M}_1$ and $\mathcal{M}_3$, for each of the datasets considered for this work with a $k_\mathrm{max.}=0.2\ h\ \mathrm{Mpc}^{-1}$. The grey dashed lines indicate two limits of the Jeffreys scale \citep{jeffreys_theory_1998}; any models with a Bayes factor greater than $\sim2.5$ have definite evidence that the sub-model is preferred, and any models with a Bayes factor greater than $\sim5$ have very strong evidence that the sub-model is preferred.}
    \label{fig:model_comp_pybird_k3}
\end{figure}

The first sub-model we consider ($\mathcal{M}_1$) is that of the fiducial setup; with $c_4$, $c_{r,2}$, and $c_\mathrm{mono.}$ all set to zero. Figure \ref{fig:model_comp_pybird_k3} shows the natural log of the Bayes factor $\ln{(B_i)}$ resulting from analysis of the \textsc{PyBird} mocks with $k_\mathrm{max.}=0.2\ h \ \mathrm{Mpc}^{-1}$ and the full likelihood. With $\ln{(B_i)}$ given by
\begin{equation}
    \ln{(B_i)}=\ln{\mathcal{Z}(\mathcal{M}_i)} - \ln{\mathcal{Z}(\mathcal{M}_0)}\ .
\end{equation}
In the above equation, $\mathcal{Z}(\mathcal{M}_0)$ is the evidence calculated for the full model, and $\mathcal{Z}(\mathcal{M}_i)$ is the evidence calculated for the sub-model being tested. We can see that although $\ln{(B_i)}$ is positive for all data samples, indicating that the sub-model is preferred, the preference is weak for all samples apart from the two eBOSS-like samples.\\

\begin{figure}
	\includegraphics[width=\linewidth]{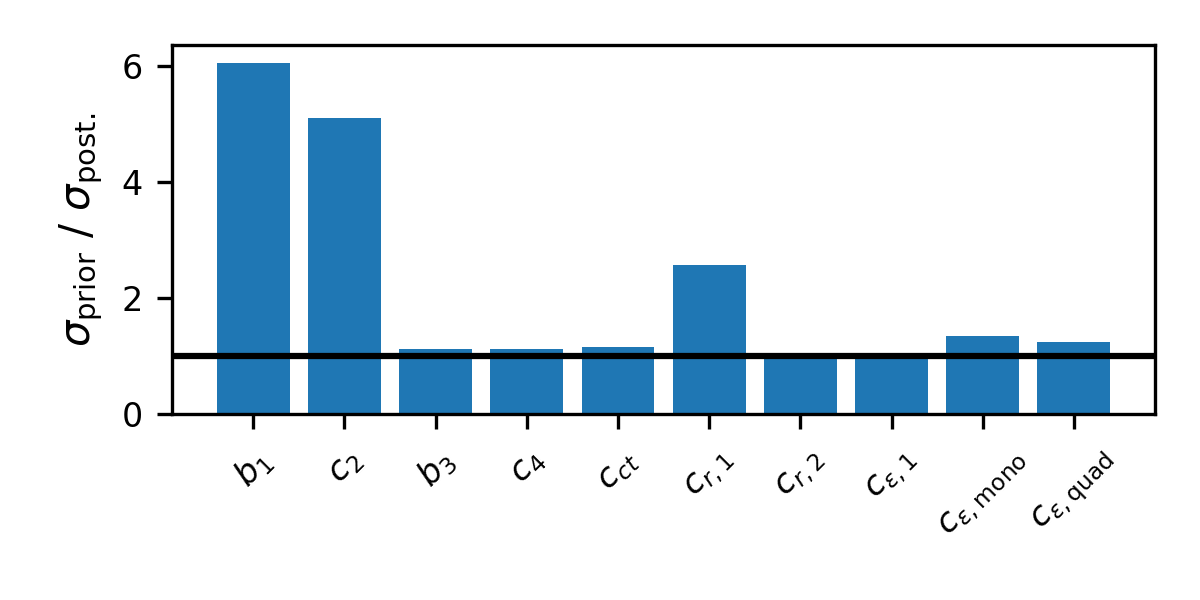}
    \caption{Ratio of the prior standard deviation to the posterior standard deviation for the marginalised 1D posteriors resulting from analysis of the BOSSz3 NGC \textsc{PyBird} mock with $k_\mathrm{max.}=0.2\ h\ \mathrm{Mpc}^{-1}$ and the full likelihood. The black solid line indicates unity.}
    \label{fig:constr_beyond_pri}
\end{figure}

The next sub-model we consider ($\mathcal{M}_3$) is chosen by observing the level constraint beyond the prior for each of the bias parameters and counterterms when analysing the \textsc{PyBird} mocks with $\mathcal{M}_0$ and $k_\mathrm{max.}=0.2\ h\ \mathrm{Mpc}^{-1}$ and the full likelihood. Figure \ref{fig:constr_beyond_pri} shows the ratio of the prior standard deviation to the 1D marginalised posterior standard deviation for each bias parameter and counterterm. We can see that the only parameters to have a significant constraint beyond the prior (ratio > 1) are $b_1$, $c_2$, and $c_{r,1}$. As such, we define sub-model $\mathcal{M}_3$ to be that with $b_1$, $c_2$, and $c_{r,1}$ as the only free nuisance parameters, and all others fixed to zero. The results of Figure \ref{fig:constr_beyond_pri} are clearly prior dependent; a reduction in the prior width for $c_{r,1}$ will result in the ratio in Figure \ref{fig:constr_beyond_pri} being lower for this parameter. These results represent the case in which we are limited to the classic prior defined in Table \ref{tab:bias_priors}. We calculate the Bayes factor for each sample in the same way as for sub-model $\mathcal{M}_1$. These Bayes factors are also plotted in Figure \ref{fig:model_comp_pybird_k3}. We can see that sub-model $\mathcal{M}_3$ is preferred over the full model $\mathcal{M}_0$ at a similar level to $\mathcal{M}_1$ for all the BOSS-like samples and the 6dFGS-like sample. However, the preference for sub-model $\mathcal{M}_3$ over the full model for the eBOSS-like samples is much stronger than sub-model $\mathcal{M}_1$. This stronger preference for the more restrictive sub-model $\mathcal{M}_3$ is likely because of the SNR of the eBOSS-like samples, as discussed in previous sections (shot noise leads to a worse SNR on small scales compared to other samples). As the parameters set to zero primarily impact small scales, and the small scales of the eBOSS-like samples are much noisier than the other samples, the data provides very little evidence for these parameters.\\

Table \ref{tab:model3_kmax_tension} shows the same as Table \ref{tab:model1_kmax_tension} for analyses of the \textsc{PyBird} mocks with sub-model $\mathcal{M}_3$. If we compare the results from the two tables, we can see that generally, the agreement is of a similar level or better than that from the results obtained with sub-model $\mathcal{M}_1$. For the eBOSS-like mocks, the level of agreement is significantly better, and the evolution with $k_\mathrm{max.}$ is now similar to that of the results from the BOSS-like mocks when considering $\ln{\left(10^{10}A_s\right)}$. These results show that we can reduce the parameter space significantly without biasing the constrained cosmology and, in some cases, can alleviate biases likely caused by volume effects.

\begin{table}
    \centering
    \begin{tabular}{c c c c c c c} 
    \hline
    Sample & \multicolumn{2}{|c|}{$\Omega_m$} & \multicolumn{2}{|c|}{$h$} & \multicolumn{2}{|c|}{$\ln{\left(10^{10}A_s\right)}$}\\ 
    \hline
    6dFGS & \textbf{0.41} & 0.42 & 0.09 & \textbf{0.04} & \textbf{2.19} & 2.51 \\ 
    BOSSz1 NGC & \textbf{0.24} & 0.25 & \textbf{0.09} & 0.23 & 0.66 & \textbf{0.03} \\ 
    BOSSz1 SGC & 0.46 & \textbf{0.43} & \textbf{0.05} & 0.11 & 1.29 & \textbf{0.78} \\ 
    BOSSz3 NGC & 0.22 & \textbf{0.02} & \textbf{0.06} & 0.1 & 0.63 & \textbf{0.07} \\ 
    BOSSz3 SGC & \textbf{0.35} & 0.43 & \textbf{0.02} & 0.14 & 1.16 & \textbf{0.7} \\ 
    eBOSS NGC & \textbf{0.25} & 0.39 & 0.12 & \textbf{0.04} & 0.78 & \textbf{0.72} \\ 
    eBOSS SGC & \textbf{0.4} & 0.49 & 0.14 & \textbf{0.1} & 1.24 & \textbf{0.91} \\ 
    \hline
    \end{tabular}
\caption{Same as Table \ref{tab:model1_kmax_tension} for analyses with sub-model $\mathcal{M}_3$ defined in Section \ref{subsubsec:model_comp}.}
\label{tab:model3_kmax_tension}
\end{table}

\subsubsection{Priors on Nuisance Parameters}
\label{subsubsec:priors}

The choice of prior for the nuisance parameters can have a significant impact on the constraint on the cosmological parameters \citep{carrilho_cosmology_2022, simon_consistency_2022}, however physically motivating priors on these parameters is challenging. The EFTofLSS is a perturbative model, and as such, if the contribution to the model from the loop corrections becomes too large, the model breaks down; this has led to priors on the nuisance parameters restricting values to be $\mathcal{O}(1)$. In this section, we explore using a Jeffreys prior \citep{jeffreys_theory_1998} as an alternative to the zero-centred Gaussian priors commonly used in the literature. We explore the use of a Jeffreys prior because it is non-informative. This is a desirable property as it means we are not favouring any particular region of the parameter space \textit{a priori}. \citet{hadzhiyska_cosmology_2023} shows that the use of the Jeffreys prior on nuisance parameters can resolve volume effects like those observed in the results presented in previous sections. The Jeffreys prior is defined as
\begin{equation}
    J(\theta) = \sqrt{|F(\theta)|}\ ,
\end{equation}
with $F(\theta)$ being the Fisher information matrix, which for a Gaussian likelihood with covariance independent of model parameters $\theta$ can be written as
\begin{equation}
    F_{ij}(\theta) = \frac{\partial M(\theta)}{\partial \theta_i}C^{-1}\frac{\partial M(\theta)}{\partial \theta_j}^T\ .
\end{equation}
From the equations above, we can see that partial derivatives of the model with respect to the model parameters are needed to evaluate the Jeffreys prior. These partial derivatives are trivial for the nuisance parameters that appear linearly in the model. They are simple sums of relevant kernels that are predicted by the \textsc{EFTEMU} (or \textsc{PyBird}) for a given set of cosmological parameters. For this work, we only impose the Jeffreys prior on these linearly appearing nuisance parameters. This means that volume effects related to these parameters should be mitigated. However, any volume effects related to marginalisation over the remaining nuisance parameters ($b_1$, $c_2$, and $c_4$) and the cosmological parameters will still remain. In practice, we impose hard bounds at -100 and 100 on the linear nuisance parameters in addition to the Jeffreys prior when using the Jeffreys prior with the full likelihood, and we impose additional Gaussian priors with $\sigma=200$ when using the Jeffreys prior with the marginalised likelihood. These additional priors are chosen relatively arbitrarily and are motivated by the practicalities of our inference pipeline\footnote{\textsc{pocoMC} requires prior samples as starting positions for particles. This means we must define a prior that we can sample from when using the full likelihood, hence the imposition of the hard bounds at -100 and 100.}. For the mock analyses presented below, the linearly appearing parameters are constrained well within the additional uniform prior when using the full likelihood. We also test setting $\sigma=1000$ when using the Jeffreys prior and see no significant difference when comparing to posteriors calculated with $\sigma=200$.\\

\begin{figure*}
	\includegraphics[width=0.95\linewidth]{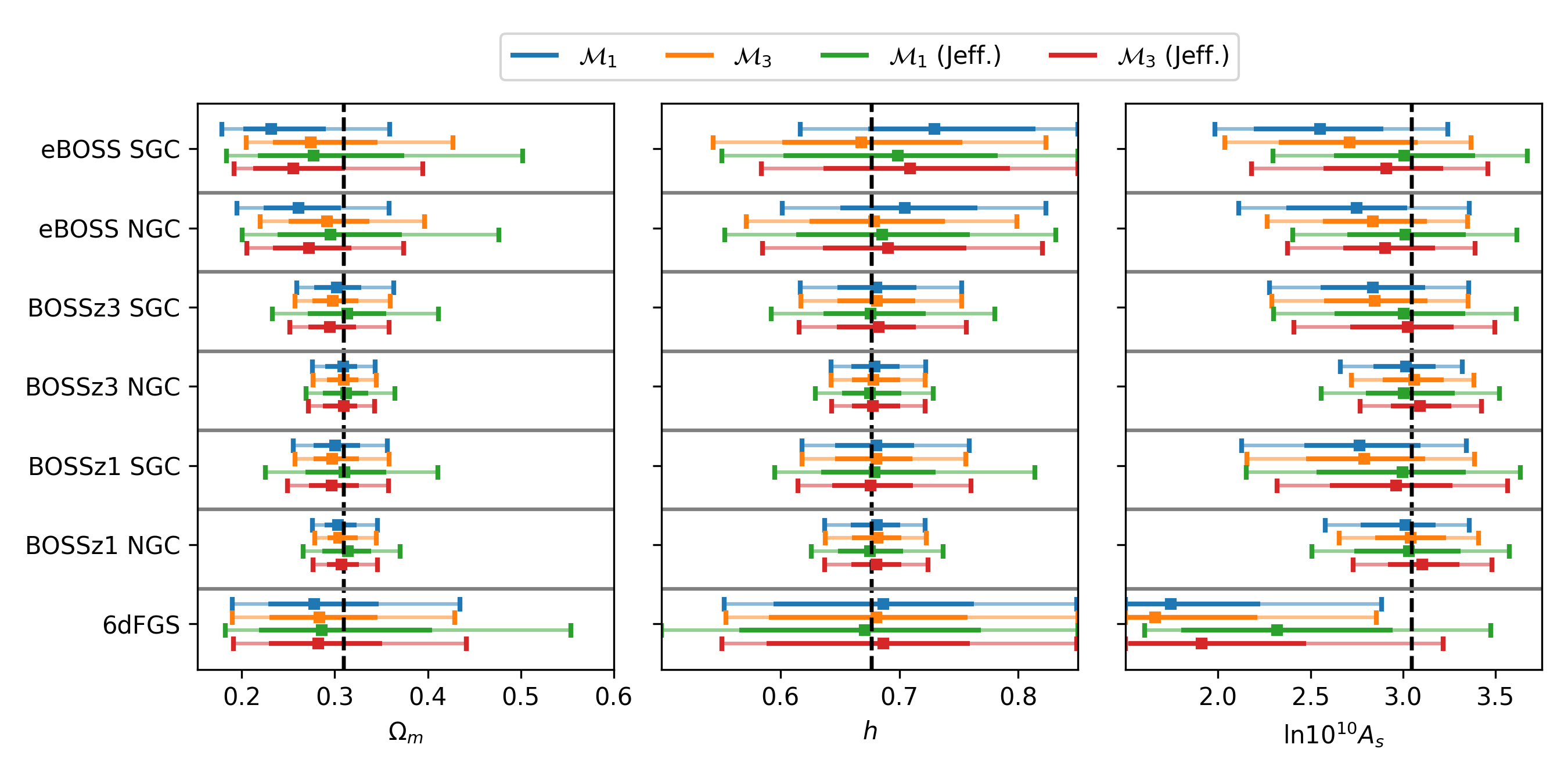}
    \caption{Same as Figure \ref{fig:model1_kmax_comp} but comparing the impact of prior choice rather than varying $k_\mathrm{max.}$. The blue and orange lines and squares show results using the classic EFTofLSS prior (defined in Table \ref{tab:bias_priors}) and sub-models $\mathcal{M}_1$ and $\mathcal{M}_3$ (defined in Section \ref{subsubsec:model_comp}) respectively. The green and red lines and squares show the results obtained using the Jeffreys prior (defined in Section \ref{subsubsec:priors}) with sub-models $\mathcal{M}_1$ and $\mathcal{M}_3$ respectively. All analyses conducted with $k_\mathrm{max.}=0.2\ h\ \mathrm{Mpc}^{-1}$.}
    \label{fig:model1_jeff_comp}
\end{figure*}

\begin{figure}
	\includegraphics[width=\columnwidth]{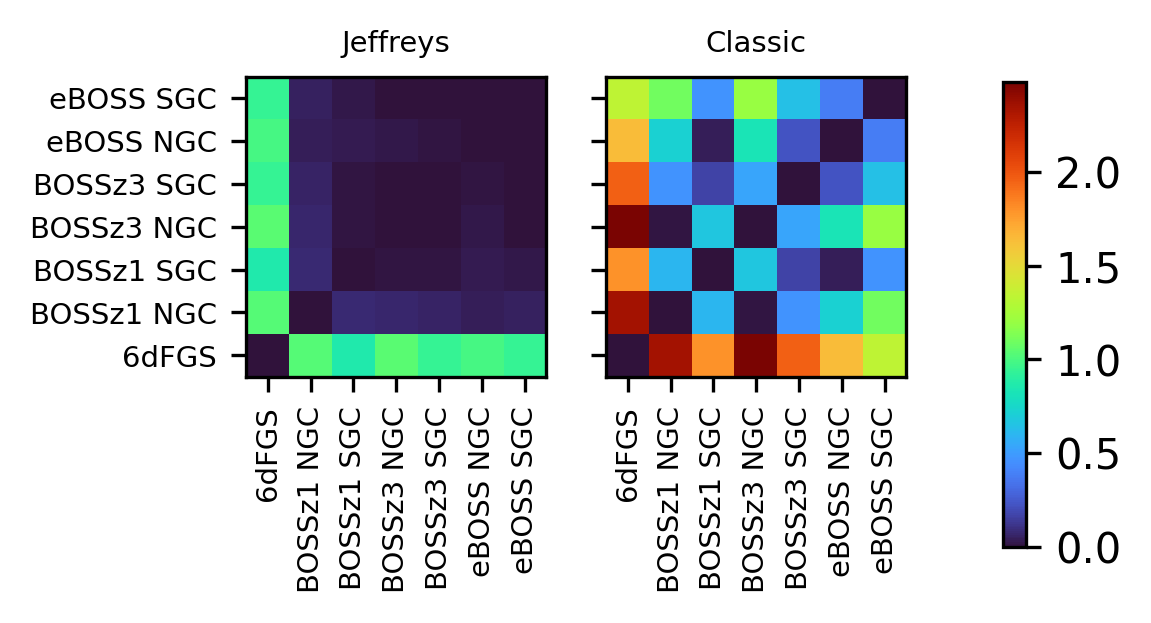}
    \caption{Matrices visualising agreement between constraints on $\ln{\left(10^{10}A_s\right)}$ resulting from analysis of different datasets with the same setup. For both panels the data was analysed with sub-model $\mathcal{M}_1$ (as defined in Section \ref{subsubsec:model_comp}) with $k_\mathrm{max.}=0.15\ h\ \mathrm{Mpc}$, and the colour indicates the magnitude of $T_{ij}$ (Equation \ref{eq:tension}). \textit{Left}: results from analysis using the Jeffreys prior defined in Section \ref{subsubsec:priors}. \textit{Right}: results from analysis with the fiducial prior.}
    \label{fig:self_tension}
\end{figure}

Figure \ref{fig:model1_jeff_comp} shows 1D marginalised posteriors for the cosmological parameters obtained from analysis of the \textsc{PyBird} mocks with sub-models $\mathcal{M}_1$ and $\mathcal{M}_3$ (defined in Section \ref{subsubsec:model_comp}), the Jeffreys prior, and the full likelihood (these setups will henceforth be referred to as JP1 and JP3, respectively). Also plotted are the results obtained with $\mathcal{M}_1$ and $\mathcal{M}_3$, the classic prior, and the full likelihood (henceforth be referred to as CP1 and CP3, respectively). We start by considering the results obtained with CP1 and JP1. We can see that for all samples, the agreement with the truth is better when using the Jeffreys prior; this is particularly noticeable for $\ln{\left(10^{10}A_s\right)}$. When using the classic prior, the $\ln{\left(10^{10}A_s\right)}$ peak posterior values shift significantly depending on the SNR of the sample. When using the Jeffreys prior, these peak posterior values are more consistently located around the true value. We expect consistency when examining the results obtained from analysis of the \textsc{PyBird} mocks as they are sample variance free. We can visualise the consistency of the results by calculating the agreement between the results obtained from each sample with Equation \ref{eq:tension} and plotting this as a matrix in Figure \ref{fig:self_tension}. We can see that for the results obtained with the Jeffreys prior, with the exception of 6dFGS, there is good agreement between the results from each other sample; however, the results obtained with the classic prior show some inconsistency. We can quantify the level of consistency by averaging the lower triangle of the matrices in Figure \ref{fig:self_tension}. This results in $0.30\sigma$ and $0.94\sigma$ for the Jeffreys prior and classic prior, respectively.\\

From Figure \ref{fig:model1_jeff_comp}, we can also see that using the Jeffreys prior results in an increase in the width of the 68\% CIs of the marginalised 1D posteriors. This should be expected, as many of the nuisance parameters converge to the prior when using the classic prior. These parameters have some degeneracy with the cosmological parameters; expanding the space that these parameters can explore inevitably leads to some degradation of the constraints on the cosmological parameters. If we examine the results obtained with JP3, we see that for the low SNR BOSS-like mocks (BOSSz1 and BOSSz3 SGC), we still have a reduction in bias in the $\ln{\left(10^{10}A_s\right)}$ constraint whilst at the same time maintaining a CI that is competitive with the classic prior. We note that for the eBOSS-like mocks, although the $\ln{\left(10^{10}A_s\right)}$ bias is reduced, it is not reduced to the same degree as with JP1. We also note that a greater bias observed in the $\Omega_m$ constraints when using the JP3 compared to the CP3.

\subsection{Joint Analyses}
\label{subsec:joint}

So far, we have considered each sample individually, which can give interesting insights into how the specifics of each sample (such as redshift and sample selection) impact the results. However, we would ultimately like to analyse multiple samples simultaneously to improve constraining power on the cosmological parameters. To do this we treat each sample as being independent, and as such define the joint likelihood as $\ln{\left[\mathcal{L}_\mathrm{joint}(\theta | \phi_\mathrm{joint})\right]} = \sum_i \ln{\left[\mathcal{L}(\theta | \phi_i)\right]}$, with $\theta$ being the shared cosmological parameters, $\phi_\mathrm{joint}$ being the complete set of nuisance parameters $\phi_\mathrm{joint}=[\phi_1, \phi_2, \ldots, \phi_n]$, and $\ln{\left[\mathcal{L}(\theta | \phi_i)\right]}$ being defined in Equation \ref{eq:likelihood}. Unless explicitly stated, the joint analyses of mocks and data measurements in this work are done with the marginalised likelihood. We exclusively use the marginalised likelihood for these kinds of analyses as the joint parameter space can become very large when considering multiple samples. The analytic marginalisation keeps the dimensionality low, thus keeping the joint analyses tractable \footnote{It is feasible to sample the parameter space for these joint analyses fully. However, we find that the number of particles for the sampler needs to be increased as suggested in the \textsc{pocoMC} documentation; \url{https://pocomc.readthedocs.io/en/latest/}. These extra particles mean extra likelihood evaluations are required for each iteration. This adds to the computational cost for each analysis that is already increased by expanding the dimensionality.}.\\

\begin{figure*}
	\includegraphics[width=\linewidth]{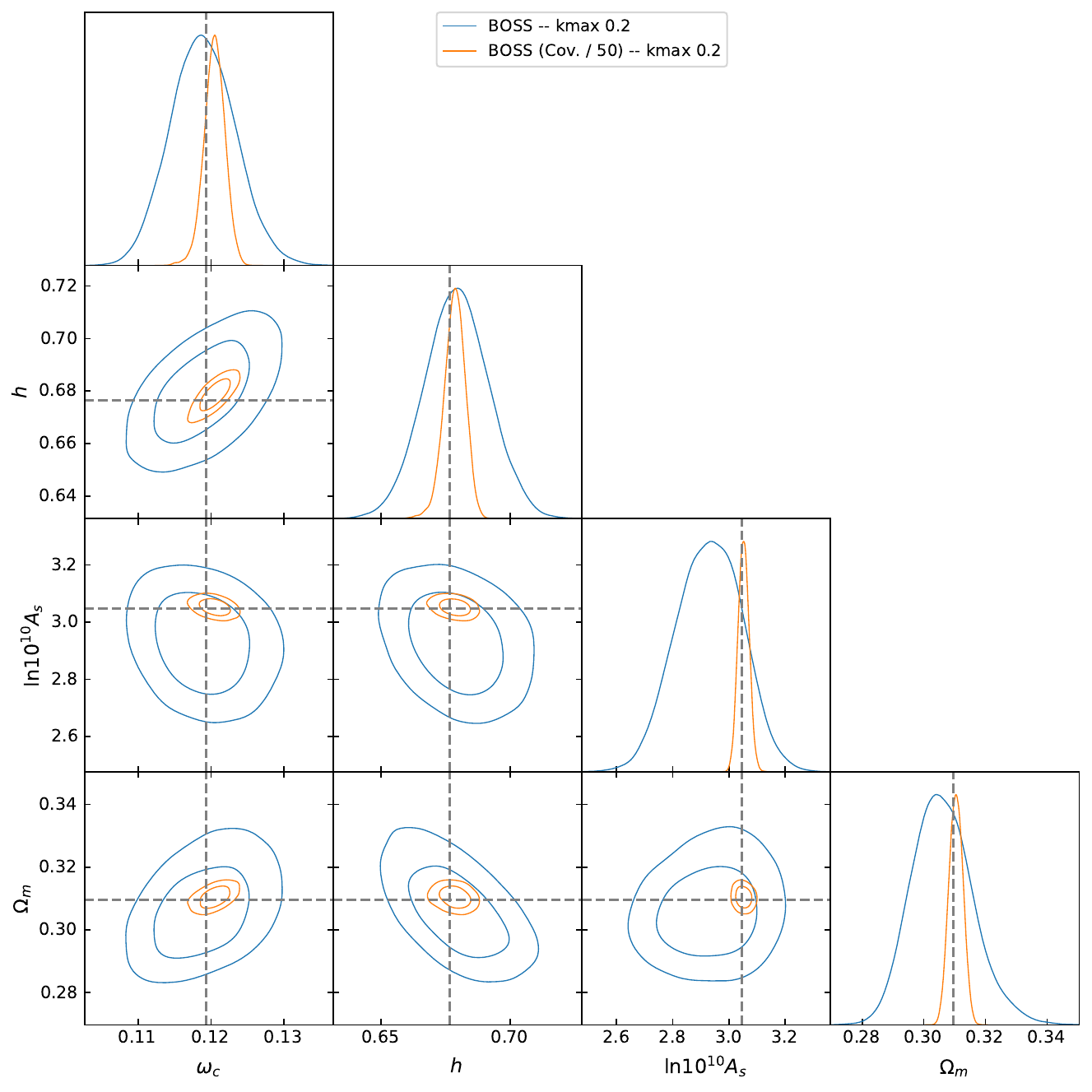}
    \caption{Same as Figure \ref{fig:corner_pybird_mock_model1} for posteriors resulting from the joint analyses of the \textsc{PyBird} mocks (discussed in Section \ref{subsec:joint}). Blue lines represent results from analysis of all BOSS-like \textsc{PyBird} mocks, orange represents the results from analysis of the BOSS-like mocks rescaled by a factor of 50. Both analyses done with sub-model $\mathcal{M}_1$ (see Section \ref{subsubsec:model_comp}), the classic style prior (see Table \ref{tab:bias_priors}), and the marginalised likelihood.}
    \label{fig:combo_BOSS_corner_mock}
\end{figure*}

Figure \ref{fig:combo_BOSS_corner_mock} shows the posterior distributions resulting from analysis of all the BOSS-like mocks (BOSSz1 NGC, BOSSz1 SGC, BOSSz3 NGC, BOSSz3 SGC) with sub-model $\mathcal{M}_1$, the classic prior, $k_\mathrm{max.}=0.2\ h\ \mathrm{Mpc}^{-1}$, and the marginalised likelihood. We note that biases can be observed in the marginalised posteriors. To verify that our joint inference pipeline does not cause these biases, we also analyse the BOSS-like mocks with the covariance rescaled by a factor of 50. These results are also plotted in Figure \ref{fig:combo_BOSS_corner_mock}. We can see that the $\sim 1\sigma$ shift from the truth when considering $\ln{\left(10^{10}A_s\right)}$ has been completely resolved. It can be seen that there is still a slight shift when considering $\omega_c$ and $h$. These biases are now more likely a result of the analysis setup, emulator error, or both rather than volume effects. We do not explore this further, as in all projections of the posterior resulting from analysis with the rescaled covariance, the truth is contained within $1\sigma$. Appendix \ref{app:pybird_comp} compares results obtained with the inference pipeline of this work with those obtained with the pipeline of \citet{ruiy_multi}.\\

\begin{figure*}
	\includegraphics[width=\linewidth]{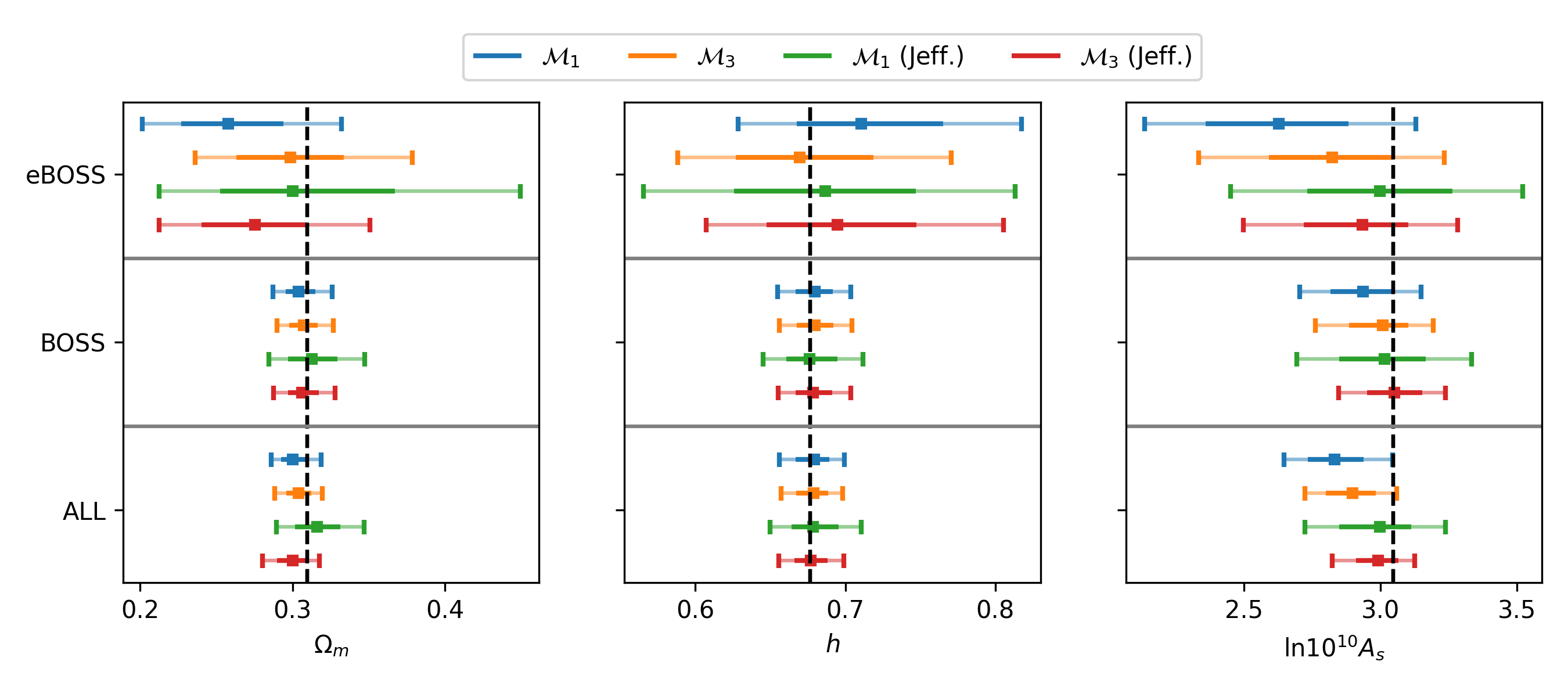}
    \caption{Same as Figure \ref{fig:model1_jeff_comp} for posteriors resulting from the joint analyses of the \textsc{PyBird} mocks (discussed in Section \ref{subsec:joint}) with various analysis setups. Blue and orange represent results from analyses with sub-models $\mathcal{M}_1$ and $\mathcal{M}_3$, and the classic style prior, respectively. Green and red show results with the Jeffreys prior defined in Section \ref{subsubsec:priors}.  $k_\mathrm{max.}=0.2\ h\ \mathrm{Mpc}^{-1}$ and the marginalised likelihood.}
    \label{fig:comb_inf_mock_various}
\end{figure*}

Figure \ref{fig:comb_inf_mock_various} summarises the marginalised 1D posteriors for the cosmological parameters of interest resulting from analyses of various combinations of the \textsc{PyBird} mocks, with various analysis setups. All analyses were conducted with $k_\mathrm{max.}=0.2\ h\ \mathrm{Mpc}^{-1}$ and the marginalised likelihood. Results obtained with sub-models $\mathcal{M}_1$ and $\mathcal{M}_3$ and the classic prior (as before referred to as CP1 and CP3) are represented with blue and orange points and lines, respectively. Results obtained with sub-models $\mathcal{M}_1$ and $\mathcal{M}_3$ and the Jeffreys prior (JP1 and JP3) are represented with green and red points and lines, respectively. Much of what can be seen from Figure \ref{fig:comb_inf_mock_various} is in line with that from Figure \ref{fig:model1_jeff_comp}. That being; when limited to the classic prior, the results obtained using CP3 are less biased than those obtained with CP1, and when considering alternative priors, the results obtained with JP1 are less biased compared to those from CP1 and CP3 at the cost of wider error bars, and although JP3 reduces the bias in the $\ln{\left(10^{10}A_s\right)}$ constraints compared to CP1 these results are more biased than those from CP3 when considering $\Omega_m$.\\

As mentioned above, the 68\%CIs are considerably wider when using JP1. This raises the question, is it even worth combing high SNR data with low SNR data if the Jeffreys prior is needed to mitigate against bias? To answer this, we look at the ratio of the 68\% CIs resulting from the joint analysis of the BOSSz1 NGC and BOSSz3 NGC \textsc{PyBird} mocks with CP1 to the 68\% CIs resulting from joint analysis of all the \textsc{PyBird} mocks with JP1. For $\Omega_m$, $h$, and $\ln{\left(10^{10}A_s\right)}$, this ratio is 0.81, 0.92, and 0.99, respectively. We can see that the use of the Jeffreys prior in JP1 has degraded the constraint in such a way that it is better to simply combine the two samples that have negligible volume effects rather than combine all samples. If we instead look at the ratio of the 68\% CIs obtained from analysis of BOSSz1 NGC and BOSSz3 NGC with CP1 to those obtained from the analysis of all the mocks with JP3, it is 1.3, 1.3, and 1.7 for $\Omega_m$, $h$, and $\ln{\left(10^{10}A_s\right)}$, respectively. In this case, there is a significant benefit from doing the joint analysis of all the samples even if the Jeffreys prior is required. It is important to note that when using JP3, we see a $\sim 1\sigma$ shift from the truth when considering $\Omega_m$. This is no worse than the bias in $\Omega_m$ seen in the results of the joint analysis of all the \textsc{PyBird} mocks with CP1 but is worse than that from the joint analysis with JP1.

\section{Main Results}
\label{sec:main_results}
In this section, we present the main results of this work; constraints on cosmological parameters from analysis of the unified power spectrum multipole measurements discussed in Section \ref{sec:data}. We repeat many of the analyses discussed in Section \ref{sec:mock_analyses}, replacing the mock multipoles with those measured from the 6dFGS, BOSS, and eBOSS redshift surveys.

\subsection{Individual Constraints}
\label{subsec:res_indiv}

\begin{figure*}
	\includegraphics[width=0.95\linewidth]{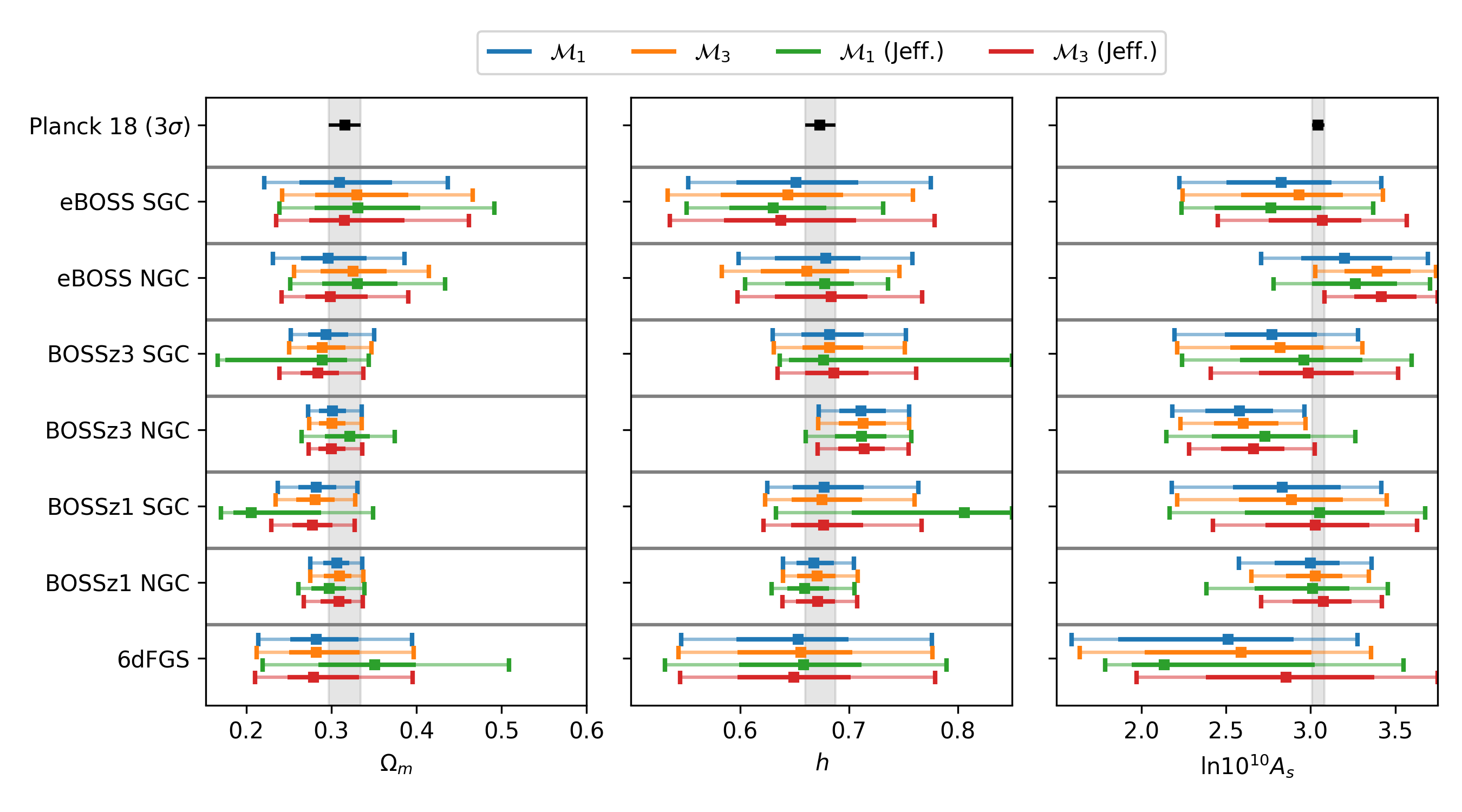}
    \caption{Same as Figure \ref{fig:model1_jeff_comp} for posteriors resulting from analysis of the multipole measurements described in Section \ref{sec:data}. Grey shaded regions and black points and lines show the peak posterior and 99\% CI from the Planck 2018 $\Lambda$CDM results. The 99\% CI has been plotted rather than 68\% (as for the other results) to aid comparison, as the Planck 68\% is much smaller than all others.}
    \label{fig:jeff_comp_obs}
\end{figure*}

We start by presenting the cosmological parameter constraints obtained via analysis of each sample individually. Figure \ref{fig:jeff_comp_obs} shows the peak posterior values and 68\% CIs for the cosmological parameters $\Omega_m$, $h$, and $\ln{\left(10^{10}A_s\right)}$ resulting from analysis of the galaxy power spectrum multipole measurements with four different setups. The first (shown with blue points and lines) being sub-model $\mathcal{M}_1$ ($c_4$, $c_{r,2}$, and $c_\mathrm{mono.}$ set to zero; see Section \ref{subsubsec:model_comp}) with the classic prior (see Table \ref{tab:bias_priors} and Section \ref{subsubsec:priors}), the next (shown with green) being sub-model $\mathcal{M}_1$ with the Jeffreys prior described in Section \ref{subsubsec:priors}, the third (shown with orange) being sub-model $\mathcal{M}_3$ (all nuisance parameters set to zero except $b_1$, $c_2$, and $c_{r,1}$) with the classic prior, and the last being sub-model $\mathcal{M}_3$ with the Jeffreys prior. We refer to these four setups as CP1, JP1, CP3, and JP3, respectively. The black points and lines, and grey shaded regions, show the 99\% CI of the Planck 2018 $\Lambda$CDM TT, TE, EE+low $\ell$+lowE+lensing+BAO results\footnote{The 99\% CIs have been plotted for Planck to make them more visible for comparison.}. The results shown in Figure \ref{fig:jeff_comp_obs} are also summarised in Table \ref{tab:individual_res}.\\

The first thing to note is the strange appearance of the CIs resulting from the analyses of BOSSz1 and BOSSz3 SGC with JP1. The marginalised 1D posteriors on $\Omega_m$ and $h$ are multimodal in these cases. The second modes of these distributions correspond to chain samples with extreme nuisance parameters. This could indicate a breakdown of the model. Further discussion on these results can be found in Appendix \ref{app:perturb_cond}.
With the exception of these results, we see good agreement between the results obtained with both sub-models and prior choices. Each given sample and parameter has agreement within $1\sigma$ for all analysis setups. However, we do note that although $<1\sigma$, there are more differences between the analysis setups when considering $\ln{\left(10^{10}A_s\right)}$.\\

Table \ref{tab:res_indiv_ten} quantifies the average level of agreement\footnote{$N_\sigma$ is calculated for each sample and the comparison via Equation \ref{eq:tension}, then these values are averaged.} between the results presented in Figure \ref{fig:jeff_comp_obs} and the Planck 2018 results. When we compare the level of agreement between the results obtained with CP1 and CP3 and the Planck 2018 results, we find that they are similar for both setups. For $\Omega_m$ and $h$ there is very little difference between the results obtained with the two setups for a majority of the samples. When considering the results from the eBOSS samples, we see more differences in the $\Omega_m$ and $h$ constraints when comparing the two setups. However, as there is a shift from an $\Omega_m$ that is lower than that from Planck 2018 to one that is higher, the average level of agreement does not change significantly. As mentioned above, the differences between the results obtained with CP1 and CP3 are clearer when considering $\ln{\left(10^{10}A_s\right)}$. For a majority of the samples, there is a shift in the peak posterior $\ln{\left(10^{10}A_s\right)}$ value towards the Planck result. This is combined with an $\sim10\%$ reduction in the width of the 68\% CIs. However, Table \ref{tab:res_indiv_ten} shows a similar level of agreement when using CP1 and CP3. This is because of the results from the eBOSS QSO NGC analysis. We can see from Figure \ref{fig:jeff_comp_obs} that the $\ln{\left(10^{10}A_s\right)}$ posterior when using CP1 is higher than that from Planck 2018. When using CP3, this shifts to even higher values. If we exclude this result, the average level of agreement in the $\ln{\left(10^{10}A_s\right)}$ constraints between the results obtained with CP1 and CP3 and the Planck results is now 1.06 and 0.860, respectively. We now consider the average level of agreement between the results obtained with JP1 and JP3 and the Planck 2018 results. We see that there is better agreement with the Planck 2018 $\ln{\left(10^{10}A_s\right)}$ constraint when compared to results obtained with CP1 for a majority of samples. As discussed in previous sections, using J1 widens the 68\% CIs. Some of the improvement in agreement with the Planck results will be because of this. However, the shifts in the peak posterior values towards the Planck 2018 results that can be seen in Figure \ref{fig:jeff_comp_obs} will also result in better agreement.\\

\begin{table}
    \centering
    \begin{tabular}{ c c c c } 
        \hline
        Comparison & $\Omega_m$ & $h$ & $\ln{\left(10^{10}A_s\right)}$ \\
        \hline
        \multirow{3}{8em}{Planck 2018} & $0.676$ & $0.499$ & $0.993$ \vspace{0.1cm} \\ 
        & $0.685$ & $0.504$ & $0.996$ \vspace{0.1cm} \\ 
        & $0.636$ & $0.687$ & $0.629$ \vspace{0.1cm} \\ 
        & $0.722$ & $0.506$ & $0.756$ \vspace{0.1cm} \\
        \hline
    \end{tabular}
\caption{The average level of agreement between the 1D marginalised posteriors resulting from the analyses (described in Section \ref{subsec:res_indiv}) of the unified multipole measurements and the Planck 2018 results. Each row corresponds to results obtained with different analysis setups. From top to bottom, those are: sub-model $\mathcal{M}_1$ with the classic prior, sub-model $\mathcal{M}_3$ with the classic prior, sub-model $\mathcal{M}_1$ with the Jeffreys prior, and model $\mathcal{M}_3$ with the Jeffreys prior.}
\label{tab:res_indiv_ten}
\end{table}

These results show that reducing model complexity going from CP1 to CP3 does not induce any statistically significant bias when considering analyses of the same sample with the two sub-models. They also show that using the reduced sub-model results in $\ln{\left(10^{10}A_s\right)}$ peak posterior values that are closer to that from Planck 2018 for the majority of data samples. We can also see that using the Jeffreys prior allows us to obtain results consistent with those obtained with the classic prior whilst being more agnostic to the form of the nuisance parameter prior. The Jeffreys prior can also increase the level of agreement with CMB results for $\ln{\left(10^{10}A_s\right)}$. However, this can come with the possible probing of unphysical regions of the parameter space.

\subsection{Joint Constraints}
\label{subsec:joint_res}

\begin{figure*}
	\includegraphics[width=\linewidth]{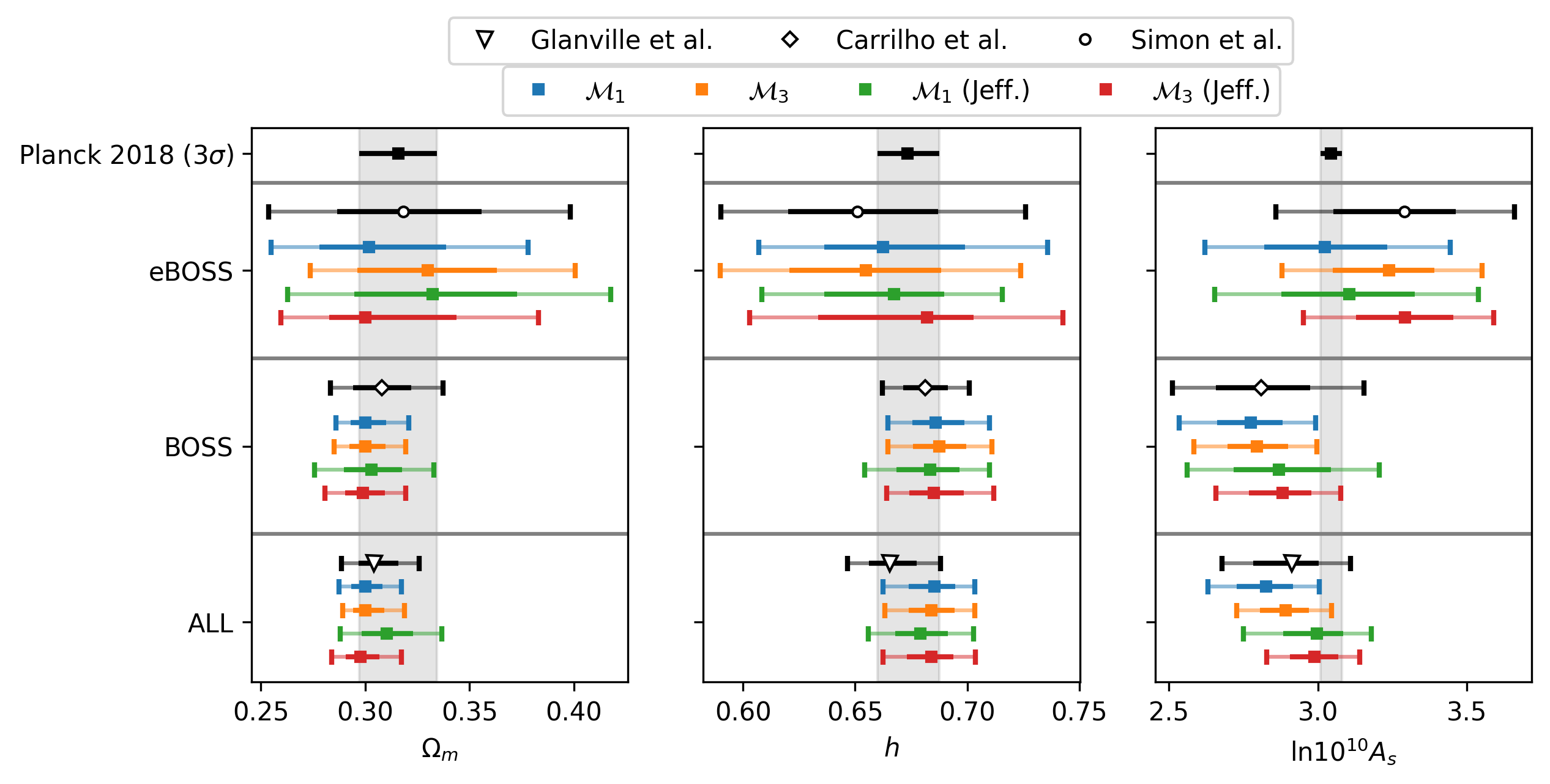}
    \caption{Same as Figure \ref{fig:comb_inf_mock_various} for the analyses of the mock multipole measurements discussed in Section \ref{sec:data}. In addition to the results from this work, the Planck 2018 results and results from \citet{carrilho_cosmology_2022, simon_cosmological_2022, glanville_full-shape_2022} are plotted for comparison. As with Figure \ref{fig:jeff_comp_obs} the 99\% CI of the Planck results have been plotted to aid comparison. All results from this work were obtained with $k_\mathrm{max.}=0.2\ h\ \mathrm{Mpc}^{-1}$ and the marginalised likelihood.}
    \label{fig:comb_inf_obs_various}
\end{figure*}

Figure \ref{fig:comb_inf_obs_various} is the same as Figure \ref{fig:jeff_comp_obs} but summarises the 1D marginalised posterior distributions on the cosmological parameters of interest resulting from joint analyses of the unified multipole measurements. These results are also summarised in \ref{tab:combo_res}. Also plotted are the Planck 2018 results and relevant results from \citet{carrilho_cosmology_2022, simon_cosmological_2022, glanville_full-shape_2022}. These works use the EFTofLSS to constrain $\Lambda$CDM parameters from analysis of galaxy power spectrum multipoles measured from different datasets. \citet{simon_cosmological_2022} uses \textsc{PyBird} to analyse the same eBOSS QSO multipoles used for this work. As with sub-model $\mathcal{M}_1$ of this work $c_4$, $c_{r,2}$, and $c_\mathrm{mono.}$ are fixed to zero. \citet{glanville_full-shape_2022} uses \textsc{PyBird} to perform joint analysis of 6dFGS, BOSS, eBOSS QSO multipole measurements. The BOSS samples used in \citet{glanville_full-shape_2022} are slightly different from those used in this work; we refer the reader to Table 1 in \citet{glanville_full-shape_2022} for details. The analysis of \citet{glanville_full-shape_2022} also differs in that the hexadecapole $P_4(k)$ is included in the data vector in addition to $P_0(k)$ and $P_2(k)$. Additionally, fewer nuisance parameters are fixed to zero than in either of the sub-models of this work. \citet{glanville_full-shape_2022} only fixes $c_4$ to zero. \citet{carrilho_cosmology_2022} uses an independent modelling pipeline for the EFTofLSS to analyse BOSS multipole measurements. Again the BOSS measurements used in \citet{carrilho_cosmology_2022} are slightly different from those used in this work; we refer the reader to Section 2.1 of \citet{carrilho_cosmology_2022} for details. The \citet{carrilho_cosmology_2022} analysis also differs in that $n_s$ is free, and the data vector includes $P_4(k)$. The form of the nuisance parameters in the \citet{carrilho_cosmology_2022} pipeline differs from that of this work \citep[for details, see Section 2.2.3 of][]{carrilho_cosmology_2022}. None of these nuisance parameters are fixed in the \citet{carrilho_cosmology_2022} analysis. Each of the three joint analysis results presented in Figure \ref{fig:comb_inf_obs_various} approximates one of the EFTofLSS works above in the sense that the same kind of data is used. The eBOSS analysis is comparable to \citet{simon_cosmological_2022}, the BOSS analysis is comparable to \citet{carrilho_cosmology_2022}, and the ALL analysis is comparable to \citet{glanville_full-shape_2022}.\\

Table \ref{tab:ten_joint_res} quantifies the level of agreement between the results of the joint analyses presented in Figure \ref{fig:comb_inf_obs_various} and the EFTofLSS literature results and the Planck 2018 $\Lambda$CDM results. We first note the good agreement between the results of each joint analysis with their respective EFTofLSS literature results. With the exception of the constraint on $h$ from the ALL analysis, we see the results of this work agree with the literature results within $\lesssim1\sigma$. The results of the joint eBOSS analysis show a more significant dependence on the analysis setup for all parameters compared to the BOSS and ALL analyses. Unlike with the analyses of the \textsc{PyBird} mocks, it is more difficult to determine if these shifts in the results are because of volume effects (resulting from a given analysis setup), sample variance, or errors in the modelling. From the mock analysis results presented in Figure \ref{fig:comb_inf_mock_various}, we see a slight shift towards the truth when using CP3. From Figure \ref{fig:comb_inf_obs_various}, we see that using CP3 shifts the results towards those of \citet{simon_cosmological_2022}. However, this shift is away from the other EFTofLSS literature results and the Planck 2018 results. If we look again at Figure \ref{fig:comb_inf_mock_various}, we see that using JP1 shifts the $\ln{\left(10^{10}A_s\right)}$ results even closer to the truth. Comparing to the equivalent result in Figure \ref{fig:comb_inf_obs_various}, we see that using JP1 shifts the $\ln{\left(10^{10}A_s\right)}$ back toward the results obtained with CP1. We note that the $\tilde{A}$ posteriors obtained with both sub-models agree very well with each other and with those from \citet{simon_cosmological_2022}. The linear bias values obtained with CP1 are $2.4\pm0.3$ and $2.3\pm0.3$ for the NGC and SGC, respectively. This aligns with the linear bias obtained via analysis of the eBOSS QSO samples with non-EFTofLSS models \citep{hou_completed_2020}. These linear bias values are significantly lower when using sub-model CP3 at $2.1\pm0.2$ for both the NGC and SGC.\\

\begin{table}
    \centering
    \begin{tabular}{c c c c c c c} 
    \hline
    Sample & \multicolumn{2}{|c|}{$\Omega_m$} & \multicolumn{2}{|c|}{$h$} & \multicolumn{2}{|c|}{$\ln{\left(10^{10}A_s\right)}$}\\ 
    \hline
    \multirow{3}{4em}{eBOSS} & $0.336$ & $0.372$ & $0.259$ & $0.293$ & $0.841$ & $0.103$ \vspace{0.1cm} \\ 
    & $0.234$ & $0.412$ & $0.077$ & $0.544$ & $0.183$ & $1.024$ \vspace{0.1cm} \\ 
    & $0.265$ & $0.43$ & $0.346$ & $0.257$ & $0.57$ & $0.269$ \vspace{0.1cm} \\ 
    & $0.336$ & $0.356$ & $0.515$ & $0.181$ & $0.009$ & $1.501$ \vspace{0.1cm} \\
    \hline
    \multirow{3}{4em}{BOSS} & $0.464$ & $1.216$ & $0.333$ & $1.066$ & $0.189$ & $2.523$ \vspace{0.1cm} \\ 
    & $0.471$ & $1.249$ & $0.404$ & $1.086$ & $0.081$ & $2.375$ \vspace{0.1cm} \\ 
    & $0.252$ & $0.768$ & $0.133$ & $0.634$ & $0.265$ & $0.996$ \vspace{0.1cm} \\ 
    & $0.52$ & $1.258$ & $0.258$ & $0.959$ & $0.355$ & $1.664$ \vspace{0.1cm} \\
    \hline
    \multirow{3}{4em}{ALL} & $0.361$ & $1.349$ & $1.199$ & $0.944$ & $0.544$ & $2.412$ \vspace{0.1cm} \\ 
    & $0.346$ & $1.297$ & $1.179$ & $0.922$ & $0.134$ & $1.908$ \vspace{0.1cm} \\ 
    & $0.368$ & $0.374$ & $0.827$ & $0.454$ & $0.58$ & $0.539$ \vspace{0.1cm} \\ 
    & $0.537$ & $1.467$ & $1.135$ & $0.861$ & $0.626$ & $0.695$ \vspace{0.1cm} \\
    \hline
    \end{tabular}
\caption{The level of agreement between the marginalised 1D posteriors on the cosmological parameters of interest resulting from the analyses described in Section \ref{subsec:joint_res}, and the Planck 2018 results and appropriate EFTofLSS literature results. For each sample, there are four rows; these correspond to results with different analysis setups. From top to bottom, they are: sub-model $\mathcal{M}_1$ with the classic prior, sub-model $\mathcal{M}_3$ with the classic prior, sub-model $\mathcal{M}_1$ with the Jeffreys prior, and sub-model $\mathcal{M}_3$ with the Jeffreys prior. Each cosmological parameter has two columns. The left column of each corresponds to the comparison with the appropriate EFTofLSS literature results: for ALL this is \citet{glanville_full-shape_2022}, \citet{carrilho_cosmology_2022} for BOSS, \citet{simon_cosmological_2022} for eBOSS. The right column shows the comparison with Planck 2018.}
\label{tab:ten_joint_res}
\end{table}

The results of the BOSS and ALL analyses show less dramatic shifts in the parameters compared to those from the eBOSS analysis and behave more like the results of the mock analyses. We see that for $\Omega_m$ and $h$, there is very little difference between the analysis setups. There is slightly better agreement with the EFTofLSS literature results and Planck 2018 for these parameters when using JP1 for both the BOSS and the ALL analysis. This results from the increased width of the 68\% CI in addition to a slight shift in the peak posterior values. The width of the 68\% and 95\% CIs appear wider from the \citet{carrilho_cosmology_2022} results than those from this work. This is most likely a result of the differences in the analysis setup mentioned above; for example, allowing $n_s$ to vary. \citet{glanville_full-shape_2022} shows an increase in the CIs of all relevant cosmological parameters when including $n_s$ as a free parameter. When we examine $\ln{\left(10^{10}A_s\right)}$, we observe that the BOSS and ALL joint analyses with CP1 display a level of agreement with the Planck 2018 results that is at the $\sim2.5\sigma$ level. For the results from CP3, this is at the $\sim2.4\sigma$ and $\sim1.9\sigma$ levels for the BOSS and ALL analyses, respectively. The results from JP1 improve the level of agreement with the Planck 2018 results for both the BOSS and ALL joint analyses to $<1\sigma$. The results from JP3 also show improved agreement with the Planck 2018 results. However, this is still $>1\sigma$ for the results from the BOSS analysis. Although the peak posterior agrees with that of the JP1 analysis, the 68\% CI is tighter and results in a $>1\sigma$ difference.\\

\section{Conclusions}
\label{sec:conclusions}
We have presented results from multiple cosmological inference analyses of mock galaxy power spectrum multipoles designed to determine how choices about the analysis setup impact the inferred cosmological parameters. To minimise the computational cost of these mock analyses, we use the neural-network-based \textsc{EFTEMU} to predict the power spectrum multiples. The training procedure of the \textsc{EFTEMU} has been improved beyond that in \citet{donald-mccann_matryoshka_2022} to allow for accurate predictions to be made on a much larger cosmological prior space. The main analysis setup choices we explore are the choice of prior on the nuisance parameters of the EFTofLSS model and which parameters to include in our analyses. The classic EFTofLSS prior takes the form of zero-centred Gaussian distributions with narrow widths on the majority of the nuisance parameters. We compare the Bayesian evidence calculated from analyses of the mock multipoles with different sets of nuisance parameters fixed at zero and the classic prior. Fixing different sets of nuisance parameters to zero results in different EFTofLSS sub-models. The first sub-model we consider ($\mathcal{M}_1$) is constructed by fixing the parameters $c_4$, $c_{r,2}$, and $c_\mathrm{mono.}$ to zero. This is a typical choice in the EFTofLSS literature. The next sub-model we consider ($\mathcal{M}_3$) is constructed by fixing all nuisance parameters but $b_1$, $c_2$, and $c_{r,1}$ to zero. There is a significant preference for sub-model $\mathcal{M}_3$ over $\mathcal{M}_1$ for the eBOSS-like mocks constructed for this work. The results of the mock analyses show less bias in the inferred cosmology when using sub-model $\mathcal{M}_3$ instead of $\mathcal{M}_1$ to analyse the eBOSS-like mocks with the classic prior.\\

The classic prior is broadly motivated by the idea of keeping the nuisance parameters small in order for the EFTofLSS model to remain perturbative. However, many of the parameters are very weakly constrained and introduce volume effects that bias the cosmological parameters upon marginalisation. We explore the use of a Jeffreys prior, a non-informative prior that can mitigate against these volume effects. Results from mock analyses with the Jeffreys prior show a significant reduction in bias in the $\ln{\left(10^{10}A_s\right)}$ constraint. Considering the joint analysis of all the mocks, we find the shift from the true $\ln{\left(10^{10}A_s\right)}$ is reduced from $2.0\sigma$ to $0.42\sigma$ comparing results obtained with sub-model $\mathcal{M}_1$ and the classic prior and Jeffreys prior respectively. The use of the Jeffreys prior comes at the cost of widening the marginalised posteriors on the cosmological parameters. This comes as a consequence of allowing the nuisance parameters to explore a much larger prior volume. For example, we find that the ratio of the width of the 68\% CI to the peak posterior value for the $\Omega_m$ marginalised posterior is $\sim7.9\%$ when analysing only the mocks with negligible volume effects, with the classic prior. If we instead analyse all of the mocks with the Jeffreys prior, this ratio is $\sim9.5\%$. This represents a $\sim20\%$ weakening of the constraint on $\Omega_m$ even though much more data has been used. We can reduce this degradation of the constraint by using the Jeffreys prior with sub-model $\mathcal{M}_3$. The more restrictive parameter space leads to a less significant widening of the 1D marginalised posteriors for the cosmological parameters when combing $\mathcal{M}_1$ with the Jeffreys prior. If we compute the ratio again, it is now $\sim6.4\%$, representing a $\sim20\%$ tightening of the constraint. Now we again see a benefit from analysing all the data, including those samples that are susceptible to volume effects.\\

From the results of the mock analyses, we expect that when the analysis setup uses sub-model $\mathcal{M}_3$ with the classic prior, we see better agreement with the truth compared to sub-model $\mathcal{M}_1$ for the eBOSS samples, and similar levels of agreement for all other samples. We also expect that when using the Jeffreys prior with both sub-models, we will see better agreement with the truth compared to analysis with sub-model $\mathcal{M}_1$ and the classic prior. Upon joint analysis of the unified multipole measurements provided in \citet{beutler_unified_2021}, we find that analysis with sub-model $\mathcal{M}_3$ and the classic prior leads to better agreement with Planck 2018 LCDM results compared to results from the same analysis with $\mathcal{M}_1$. The level of agreement is improved from $2.4\sigma$ to $1.9\sigma$ for $\ln{\left(10^{10}A_s\right)}$. Analysing all of the multipoles with the Jeffreys prior and both sub-models leads to better levels of agreement again. This is now $0.54\sigma$ and $0.70\sigma$ for $\mathcal{M}_1$ and $\mathcal{M}_3$ respectively. These results indicate that some of the slight tensions between results obtained via analysis with the EFTofLSS and those obtained via analysis of the CMB are a result of analysis setup.\\

When using the Jeffreys prior, the nuisance parameters can take extreme values. From analysis of the individual data samples, some cases indicate that using the Jeffreys prior allows for probing regions of the parameters space in which the EFTofLSS model is no longer valid. This presents as multimodal distributions in the cosmological parameters. As mentioned above, we also see a degradation of the cosmological parameter constraint when using the Jeffreys prior. Both of these issues can be addressed by the inclusion of the classic prior within the Jeffreys prior. This limits how extreme the nuisance parameters can be, removing potentially unphysical regions of the parameter space and preventing degradation of the cosmological parameter constraints. This is explored in \citet{ruiy_multi}. It is shown that the resolution of volume effects can be achieved without severe degradation of the cosmological parameter constraints. We do not explore this in this work as the values of the prior widths for the nuisance parameters represent extra hyperparameters of the analysis that need to be motivated. The use of the Jeffreys prior without this additional tight prior represents an agnostic approach to the nuisance parameters, whether all regions of the parameter space are physically valid or not.

\section*{Acknowledgements}
The authors would like to thank the authors of \citet{carrilho_cosmology_2022, simon_cosmological_2022,glanville_full-shape_2022} for sharing their MCMC chains. The authors would also like to thank the ICG LSS journal club members, who have provided valuable comments and feedback during various stages of this work. JD-M and RG were supported by STFC studentships. RZ is supported by NSFC grants 11925303 and 11890691, and is also supported by the Chinese Scholarship Council (CSC) and the University of Portsmouth. KK is supported by the STFC grant ST/W001225/1. FB is a University Research Fellow and has received funding from the European Research Council (ERC) under the European Union’s Horizon 2020 research and innovation program (grant
agreement 853291). For the purpose of open access, the authors have applied a Creative Commons Attribution (CC BY) licence to any Author Accepted Manuscript version arising.

\section*{Data Availability}
Multipole measurements, covariance matrices, window function matrices, and wide-angle matrices can be found at \url{https://fbeutler.github.io/hub/deconv_paper.html}. The Planck 2018 $\Lambda$CDM chains used for comparison in this work are available at \url{https://wiki.cosmos.esa.int/planck-legacy-archive/index.php/Cosmological_Parameters}. The \textsc{PyBird} mocks, posterior samples, and inference pipeline will be made publicly available upon acceptance.

\bibliographystyle{mnras}
\bibliography{ref}

\appendix

\section{Remaining Perturbative}
\label{app:perturb_cond}
One of the main arguments for using zero-centred Gaussian priors on the parameters that appear linearly in the EFTofLSS model is that if these parameters become too large, the model is no longer perturbative. We define a simple check for the model remaining perturbative by evaluating two conditions,
\begin{equation}
    \mathrm{Condition 1}: \left|\sum_i P_{i,l}^{11}(k)b_i^{11}\right| > \left|\sum_i P_{i,l}^\mathrm{loop}(k)b_i^\mathrm{loop}\right|\ ,
\end{equation}

\begin{equation}
    \mathrm{Condition 2}: \left|\sum_i P_{i,l}^{11}(k)b_i^{11}\right| > \left|\sum_i P_{i,l}^\mathrm{ct.}(k)b_i^\mathrm{ct.}\right|\ ,
\end{equation}
with $P_{i,l}^{11}(k)$, $P_{i,l}^\mathrm{loop}(k)$, and $P_{i,l}^\mathrm{ct.}(k)$ being the kernels associated to linear, loop, and counterterm contributions respective, $b_i^{11}$, $b_i^\mathrm{loop}$, and $b_i^\mathrm{ct.}$ being the bias parameters, or counterterms, or combination associated to each kernel. If either of the above conditions fails, we say the model is no longer perturbative. Whilst finalising this work, the authors were made aware of \citet{braganca_peeking_2023}, in which a "perturbativity prior" is introduced. We note some similarity as this perturbativity prior sets a Gaussian prior on the overall loop term. However, the perturbative condition above was developed independently of \citet{braganca_peeking_2023} and is different in that a uniform probability is given to models that pass the perturbative condition.\\

\begin{figure*}
	\includegraphics[width=\linewidth]{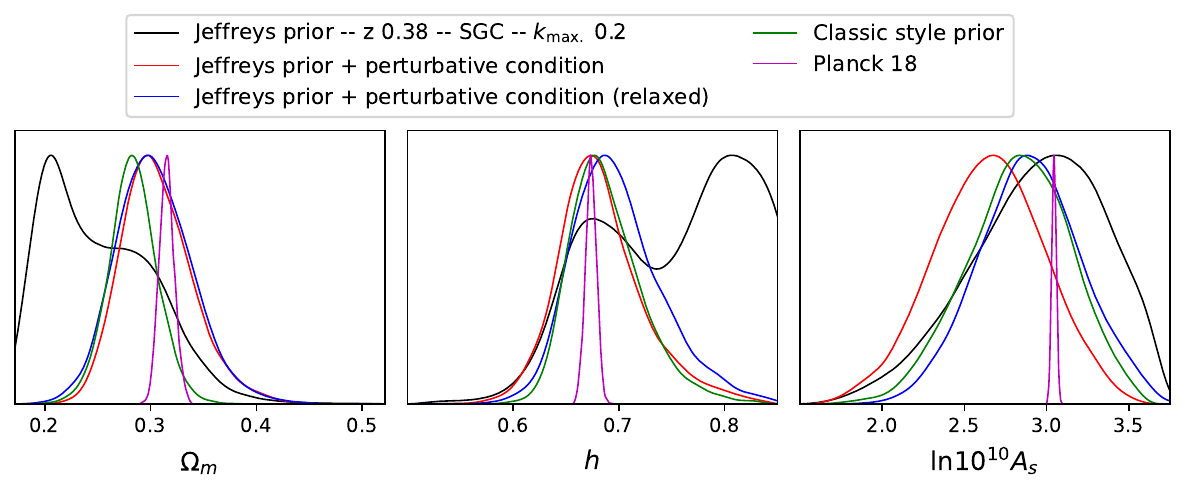}
    \caption{Marginalised 1D posteriors for the cosmological parameters of interest obtained with different analysis setups when analysing the BOSSz1 SGC multipole measurements. All analyses were conducted with $k_\mathrm{max.} = 0.2\ h\ \mathrm{Mpc}^{-1}$ and the full likelihood. The black lines represent the results obtained with the Jeffreys prior (defined in Section \ref{subsubsec:priors}), the green lines represent the results with the classic prior (defined in Table \ref{tab:bias_priors}), and the red and blues represent the results obtained with the Jeffreys prior and the perturbative conditions discussed in Appendix \ref{app:perturb_cond}. Also plotted (in purple) are the marginalised 1D posteriors for the Planck 2018 results.}
    \label{fig:perturb_cond}
\end{figure*}

Figure \ref{fig:perturb_cond} shows the marginalised 1D posteriors for the cosmological parameters of interest obtained with different analysis setups when analysing the BOSSz1 SGC multipole measurements. In black are the results obtained when using the Jeffreys prior. We note the unusual shape of the marginalised posteriors for $\Omega_m$ and $h$; there are second modes of the distributions very far from the Planck 2018 results (plotted in Figure \ref{fig:perturb_cond} in purple) and at the extremes of the prior-space. The nuisance parameters associated with the cosmological parameters of these second modes tend to have more extreme values than those in the central mode. This potentially indicates a breakdown in the EFTofLSS model in these regions. We can test this by imposing the perturbative condition defined by the equations above when sampling\footnote{In practice, this involves heavily penalising any sample that breaks this condition, rather than imposing a hard bound}. The red lines in Figure \ref{fig:perturb_cond} show the results obtained when doing this. We can see that the extreme second modes in the $\Omega_m$ and $h$ posteriors have been removed. However, this also results in a shift in $\ln{\left(10^{10}A_s\right)}$ to lower values. These results appear to show that imposing the perturbative condition when sampling negates the effects of the Jeffreys prior when it comes to resolving the volume effects. A possible cause for this is the way that the perturbative condition has been included. Imposing the condition reduces the prior volume. However, no change has been made to the Jeffreys prior. So the prior volume corrected for with the inclusion of the Jeffreys prior is not the prior volume explored. It should be noted that these results are obtained with a prior that is still more agnostic than the classic style prior. We have made no choice on how large the nuisance parameters can be. We also note that the condition defined by the equations above is by no means exact. If we use a slightly more relaxed version of the condition,

\begin{multline}
\mathrm{Condition (relaxed)}: \left|\sum_i P_{i,l}^{11}(k)b_i^{11}\right|\\ > \left|\sum_i P_{i,l}^\mathrm{ct.}(k)b_i^\mathrm{ct.}+\sum_i P_{i,l}^\mathrm{loop}(k)b_i^\mathrm{loop}\right|\ ,
\end{multline}
then we obtain the results shown with blue lines in Figure \ref{fig:perturb_cond}. In this case, the $\ln{\left(10^{10}A_s\right)}$ constraint agrees with that obtained with the classic style prior (shown with green), but again is a more agnostic prior than the classic style prior.

\section{Toy Model}
\label{app:toy}
The purpose of this appendix is to give some intuition for the volume effect discussed frequently in the main text. All of what follows is based on \citet[][see Section 2 for a detailed discussion of the concepts discussed in this appendix]{hadzhiyska_cosmology_2023}.\\

\citet{hadzhiyska_cosmology_2023} shows that the marginalised $\chi^2$ can be written as
\begin{equation}
    \chi^2_m(\Omega) \simeq \chi^2_\ast(\Omega) + \log \left\{ \det\left[\mathcal{F}_\ast(\Omega)\right] \right\} + \mathrm{const.}\ ,
    \label{eq:marg_chi}
\end{equation}
with $\chi^2_\ast(\Omega) = \chi^2(\Omega, n_\ast)$, where $\Omega$ and $n_\ast$ are the model parameters of interest and $n_\ast$ being the best-fit nuisance parameters, respectively, and $\mathcal{F}_\ast(\Omega)$ is given by
\begin{equation}
    \mathcal{F}_{\ast,ij}(\Omega) = \frac{1}{2}\frac{\partial^2\chi^2}{\partial n_i\partial n_j} \Bigr\rvert_{n_\ast} \ .
\end{equation}
The two terms in Equation \ref{eq:marg_chi} are referred to as the \textit{profile} and \textit{Laplace} terms, respectively. The Laplace term is responsible for the volume effect that induces biases in the marginalised posterior on the parameters of interest $\Omega$ when marginalising over the nuisance parameters $n$.\\

To aid in understanding, we present results from analysis with a toy model. This toy model is the same as that discussed in Section 2.3 of \citet{hadzhiyska_cosmology_2023}; it is a simple power law of the form $n x^\Omega$. We start by defining true values for $n$ and $\Omega$. For the example here, we use 50 and 1.5, respectively. We then generate 100 $x$-values as random draws from a uniform distribution $\mathcal{U}(1,5)$. We compute the model prediction for these 100 $x$-values and true parameters and associate the same uncertainty $\sigma$ to each of them (this corresponds to a covariance matrix with a constant diagonal and zero off-diagonals). This synthetic data is then analysed with the same kind of inference pipeline used for the work presented in the main text. When $\sigma=25$ the inferred $\Omega$ after marginalisation over $n$ is $1.500^{+0.031}_{-0.033}$; very good agreement with the true $\Omega$. However, when $\sigma=800$ the inferred $\Omega$ is $0.73^{+0.93}_{-0.91}$; we now have a shift in the peak posterior. This shift in the peak posterior has come from the Laplace term.\\

\begin{figure}
	\includegraphics[width=\columnwidth]{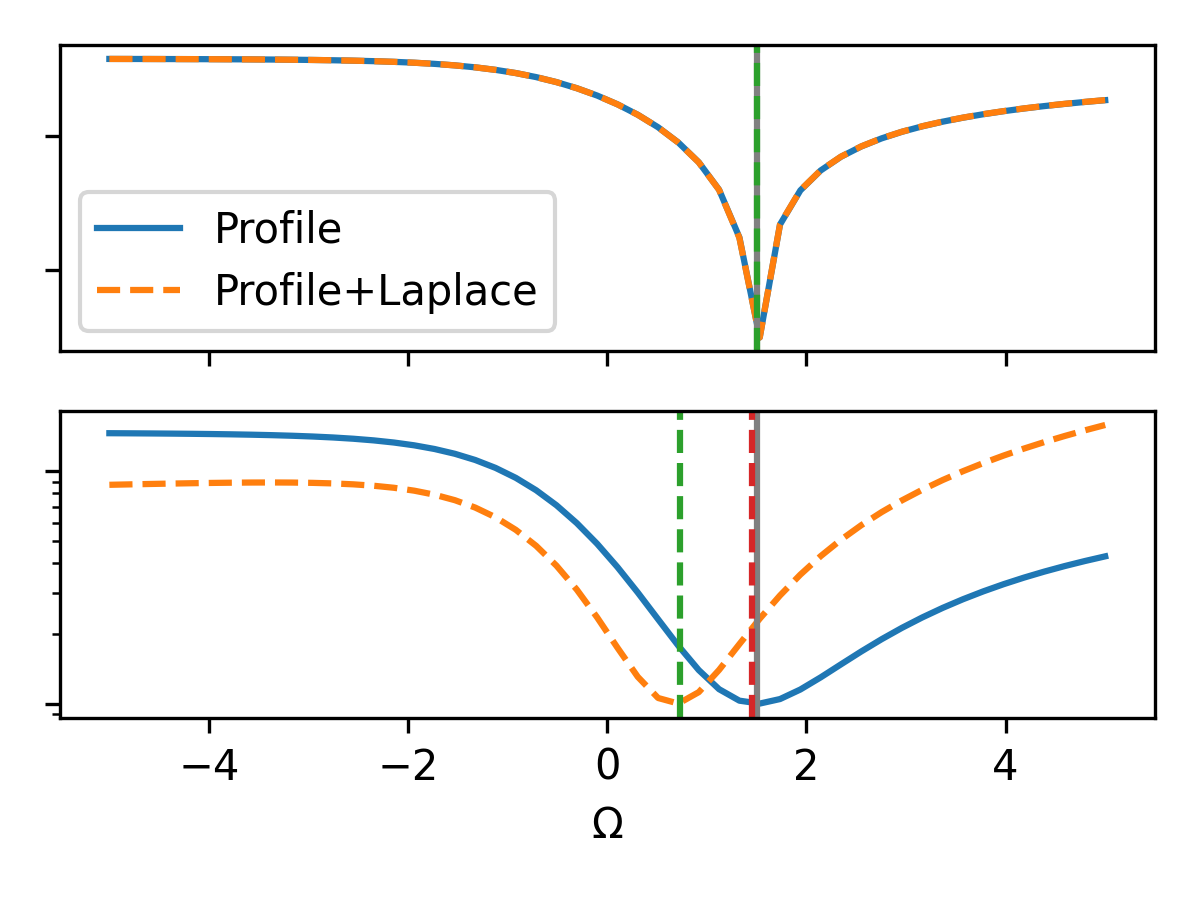}
    \caption{Comparison of the profile and Laplace terms discussed in Appendix \ref{app:toy}. The top panel shows the terms calculated for each $\Omega$ with $\sigma=25$. The bottom panel shows the same with $\sigma=800$. The grey solid lines indicate the location of the true $\Omega$. The green dashed lines indicate the peak of the marginalised posterior obtained from carrying out inference with the given value of $\sigma$. The red dashed line in the bottom panel shows the peak posterior when conducting inference with a Jeffreys prior.}
    \label{fig:toy_laplace}
\end{figure}

Figure \ref{fig:toy_laplace} compares the profile and Laplace terms for the two values of $\sigma$. The top panel shows the results with $\sigma=25$, and the bottom panel shows the results with $\sigma=800$. The blue lines show the profile term $\chi_\ast^2(\Omega)$, and the dashed orange lines show the sum of the profile term and the Laplace term $\log \left\{ \det\left[\mathcal{F}_\ast(\Omega)\right] \right\}$. The results have been normalised such that the minimum has a value of zero. The grey solid line shows the location of the true $\Omega$. We can see that when $\sigma=25$, both minima are in agreement with the truth; the Laplace term has a negligible impact when the constraining power is high. When $\sigma=800$, we see that the minimum of the profile term is still located at the truth. However, the sum of the Laplace and profile terms is shifted toward lower values of $\Omega$.

As discussed in Section \ref{subsubsec:priors},  \citet{hadzhiyska_cosmology_2023} show that a Jeffreys prior can be used to mitigate against the volume effect. Figure \ref{fig:toy_laplace} also shows the peak posterior for $\Omega$ when carrying out inference with a Jeffreys prior with a red dashed line. For this toy example, we only have one nuisance parameter $n$. As such, the Fisher matrix needed to evaluate the Jeffreys prior is a single value. Given by
\begin{equation}
    F = \frac{\partial\left(n x^\Omega\right)}{\partial n}\bm{C}^{-1}\frac{\partial\left(n x^\Omega\right)}{\partial n}^T = x^\Omega \bm{C}^{-1} \left(x^\Omega\right)^T\ . 
\end{equation}
We can see from Figure \ref{fig:toy_laplace} that the marginalised peak posterior is now in good agreement with the truth and minimum of the profile term.\\

We can relate this toy example to the work of the main text by considering the form of the toy model. We can think of $n$ as being one of the linearly appearing nuisance parameters of the EFTofLSS model and $x^\Omega$ as the kernel or combination of kernels relevant for that nuisance parameter.

\section{Comparison with \textsc{PyBird}}
\label{app:pybird_comp}
\begin{figure*}
	\includegraphics[width=\linewidth]{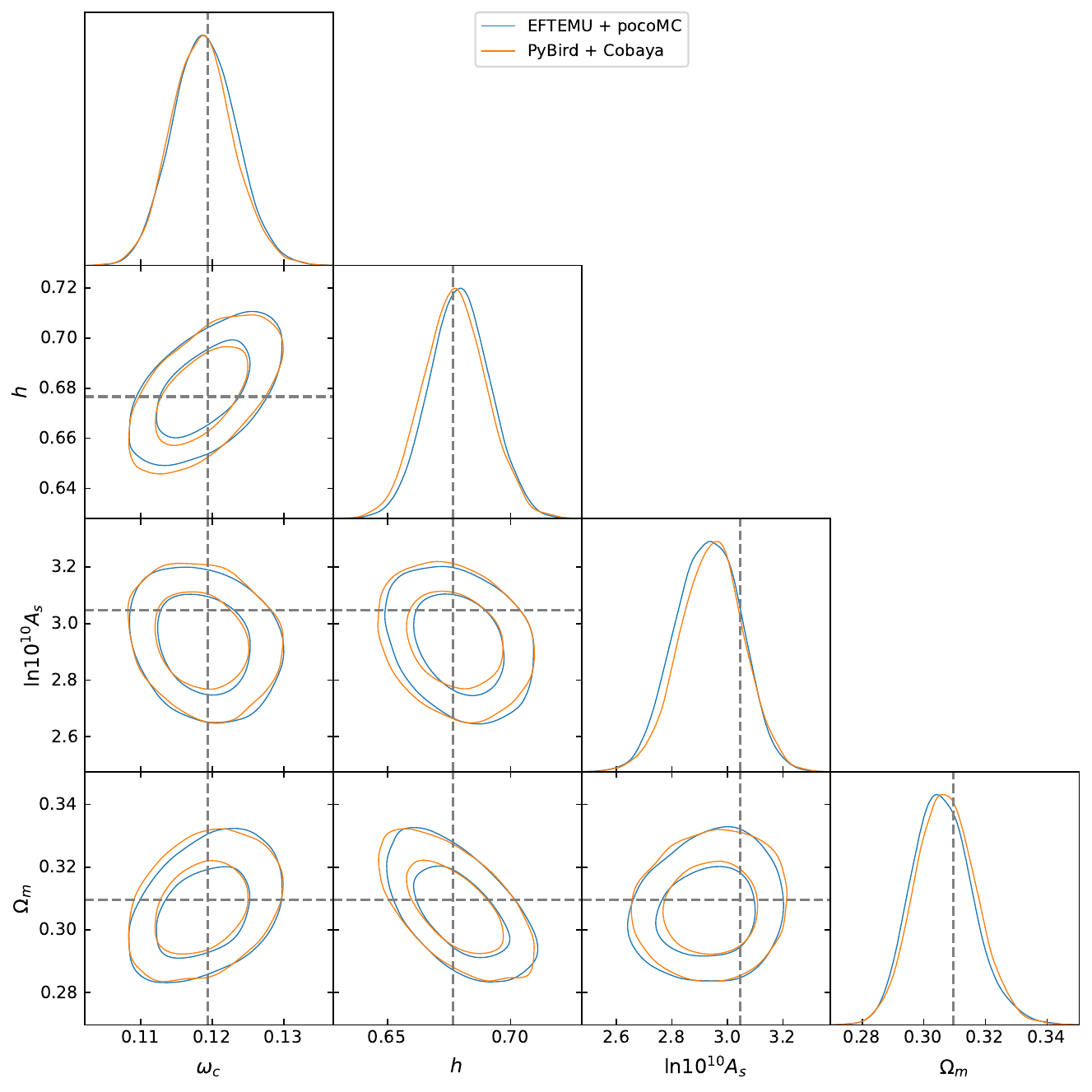}
    \caption{Same as Figure \ref{fig:corner_pybird_mock_model1} showing the results of joint analyses of the BOSS-like \textsc{PyBird} mocks. Blue lines show results obtained with the \textsc{EFTEMU} and the cosmological inference pipeline developed for this work. Orange results were obtained using \textsc{PyBird} and the inference pipeline of \citet{ruiy_multi}. Both analyses were done with the classic prior, the marginalised likelihood, and $k_\mathrm{max.}= 0.2\ h\ \mathrm{Mpc}^{-1}$.}
    \label{fig:pybird_comp}
\end{figure*}

Figure \ref{fig:pybird_comp} compares posterior distributions from the joint analysis of the BOSS-like \textsc{PyBird} mocks with two different inference pipelines. The first is the pipeline of this work, with model predictions from the \textsc{EFTEMU} and sampling with \textsc{pocoMC}. The second pipeline is a slightly modified version of that used in \citet{ruiy_multi}, with model predictions being made with \textsc{PyBird} and the sampling done with \textsc{Cobaya} \citep{torrado_cobaya_2021}. Both analyses were conducted with the marginalised likelihood, $k_\mathrm{max.}=0.2\ h\ \mathrm{Mpc}^{-1}$, sub-model $\mathcal{M}_1$, and the classic prior. The percentage difference in the posterior means is less than 0.5\% for all parameters in Figure \ref{fig:pybird_comp}. The percentage difference in the width of the 68\% CIs is, at worst, 5\%.

\section{Results Tables}
\label{app:results_tabs}
Table \ref{tab:individual_res} summarises the cosmological constraints resulting from the unified multipole measurements discussed in Section \ref{subsec:res_indiv}. Table \ref{tab:combo_res} summarises the cosmological constraints resulting from the joint analyses discussed in Section \ref{subsec:joint_res}.

\begin{table}
    \centering
    \begin{tabular}{ c c c c } 
        \hline
        Sample & $\Omega_m$ & $h$ & $\ln{\left(10^{10}A_s\right)}$ \\
        \hline
            \multirow{3}{8em}{6dFGS} & $0.283_{-0.031}^{+0.05}$ & $0.653_{-0.057}^{+0.046}$ & $2.51_{-0.65}^{+0.39}$ \vspace{0.1cm} \\ 
            & $0.283_{-0.032}^{+0.051}$ & $0.656_{-0.058}^{+0.047}$ & $2.59_{-0.57}^{+0.42}$ \vspace{0.1cm} \\ 
            & $0.351_{-0.066}^{+0.048}$ & $0.658_{-0.059}^{+0.053}$ & $2.14_{-0.19}^{+0.89}$ \vspace{0.1cm} \\ 
            & $0.279_{-0.031}^{+0.054}$ & $0.65 \pm 0.052$ & $2.86_{-0.47}^{+0.52}$ \vspace{0.1cm} \\ 
            \hline
            \multirow{3}{8em}{BOSSz1 NGC} & $0.307_{-0.016}^{+0.014}$ & $0.668_{-0.016}^{+0.018}$ & $3.0_{-0.21}^{+0.17}$ \vspace{0.1cm} \\ 
            & $0.309_{-0.018}^{+0.014}$ & $0.671_{-0.018}^{+0.017}$ & $3.03_{-0.17}^{+0.16}$ \vspace{0.1cm} \\ 
            & $0.298_{-0.021}^{+0.019}$ & $0.659_{-0.016}^{+0.022}$ & $3.01_{-0.34}^{+0.22}$ \vspace{0.1cm} \\ 
            & $0.309_{-0.022}^{+0.015}$ & $0.671_{-0.02}^{+0.016}$ & $3.08_{-0.18}^{+0.16}$ \vspace{0.1cm} \\ 
            \hline
            \multirow{3}{8em}{BOSSz1 SGC} & $0.282_{-0.021}^{+0.024}$ & $0.677_{-0.029}^{+0.036}$ & $2.83_{-0.29}^{+0.34}$ \vspace{0.1cm} \\ 
            & $0.281_{-0.022}^{+0.023}$ & $0.675_{-0.028}^{+0.037}$ & $2.89_{-0.31}^{+0.3}$ \vspace{0.1cm} \\ 
            & $0.206_{-0.021}^{+0.082}$ &  > 0.7 & $3.05_{-0.44}^{+0.39}$ \vspace{0.1cm} \\ 
            & $0.278 \pm 0.024$ & $0.677_{-0.03}^{+0.036}$ & $3.03_{-0.3}^{+0.32}$ \vspace{0.1cm} \\ 
            \hline
            \multirow{3}{8em}{BOSSz3 NGC} & $0.301 \pm 0.016$ & $0.711_{-0.02}^{+0.023}$ & $2.58 \pm 0.2$ \vspace{0.1cm} \\ 
            & $0.301_{-0.015}^{+0.016}$ & $0.713_{-0.023}^{+0.02}$ & $2.6_{-0.17}^{+0.21}$ \vspace{0.1cm} \\ 
            & $0.322_{-0.029}^{+0.024}$ & $0.712_{-0.024}^{+0.023}$ & $2.73_{-0.31}^{+0.27}$ \vspace{0.1cm} \\ 
            & $0.3_{-0.015}^{+0.017}$ & $0.714_{-0.024}^{+0.019}$ & $2.66_{-0.19}^{+0.18}$ \vspace{0.1cm} \\ 
            \hline
            \multirow{3}{8em}{BOSSz3 SGC} & $0.294_{-0.021}^{+0.026}$ & $0.682_{-0.026}^{+0.031}$ & $2.77_{-0.28}^{+0.26}$ \vspace{0.1cm} \\ 
            & $0.289_{-0.018}^{+0.027}$ & $0.683_{-0.025}^{+0.031}$ & $2.82_{-0.29}^{+0.26}$ \vspace{0.1cm} \\ 
            & $0.289_{-0.113}^{+0.029}$ & $0.677_{-0.032}^{+0.171}$ & $2.96_{-0.38}^{+0.35}$ \vspace{0.1cm} \\ 
            & $0.284_{-0.02}^{+0.025}$ & $0.686_{-0.026}^{+0.032}$ & $2.98_{-0.29}^{+0.27}$ \vspace{0.1cm} \\ 
            \hline
            \multirow{3}{8em}{eBOSS QSO NGC} & $0.296_{-0.031}^{+0.045}$ & $0.679_{-0.047}^{+0.032}$ & $3.2_{-0.26}^{+0.28}$ \vspace{0.1cm} \\ 
            & $0.326 \pm 0.039$ & $0.662_{-0.042}^{+0.039}$ & $3.39_{-0.19}^{+0.2}$ \vspace{0.1cm} \\ 
            & $0.33_{-0.041}^{+0.047}$ & $0.678_{-0.036}^{+0.027}$ & $3.26 \pm 0.25$ \vspace{0.1cm} \\ 
            & $0.299_{-0.029}^{+0.044}$ & $0.684_{-0.051}^{+0.033}$ & $3.42_{-0.16}^{+0.21}$ \vspace{0.1cm} \\ 
            \hline
            \multirow{3}{8em}{eBOSS QSO SGC} & $0.309_{-0.047}^{+0.062}$ & $0.651_{-0.054}^{+0.057}$ & $2.83_{-0.32}^{+0.3}$ \vspace{0.1cm} \\ 
            & $0.33_{-0.049}^{+0.061}$ & $0.644_{-0.062}^{+0.051}$ & $2.93_{-0.34}^{+0.26}$ \vspace{0.1cm} \\ 
            & $0.331_{-0.05}^{+0.073}$ & $0.63_{-0.04}^{+0.049}$ & $2.77_{-0.33}^{+0.3}$ \vspace{0.1cm} \\ 
            & $0.315_{-0.041}^{+0.071}$ & $0.637_{-0.052}^{+0.069}$ & $3.07_{-0.32}^{+0.23}$ \vspace{0.1cm} \\ 
            \hline
    \end{tabular}
    \caption{Peak posterior values and 68\% CIs for the cosmological parameters of interest resulting from analysis of the individual datasets considered for this work with $k_\mathrm{max.}=0.2\ h\ \mathrm{Mpc}^{-1}$ and the full likelihood. Each row is split into four. These correspond to results obtained with different analysis setups. From top to bottom, those are: sub-model $\mathcal{M}_1$ with the classic prior, sub-model $\mathcal{M}_3$ with the classic prior, sub-model $\mathcal{M}_1$ with the Jeffreys prior, and model $\mathcal{M}_3$ with the Jeffreys prior.}
    \label{tab:individual_res}
\end{table}

\begin{table}
    \centering
    \begin{tabular}{ c c c c } 
        \hline
        Sample & $\Omega_m$ & $h$ & $\ln{\left(10^{10}A_s\right)}$ \\
        \hline
\multirow{3}{8em}{ALL} & $0.3_{-0.007}^{+0.008}$ & $0.685_{-0.011}^{+0.009}$ & $2.83_{-0.1}^{+0.09}$ \vspace{0.1cm} \\ 
& $0.3_{-0.006}^{+0.009}$ & $0.684 \pm 0.01$ & $2.89_{-0.09}^{+0.08}$ \vspace{0.1cm} \\ 
& $0.31_{-0.012}^{+0.013}$ & $0.679_{-0.011}^{+0.012}$ & $3.0_{-0.11}^{+0.09}$ \vspace{0.1cm} \\ 
& $0.298_{-0.007}^{+0.009}$ & $0.684_{-0.011}^{+0.01}$ & $2.99 \pm 0.08$ \vspace{0.1cm} \\ 
\hline
\multirow{3}{8em}{BOSS} & $0.3_{-0.007}^{+0.01}$ & $0.686_{-0.01}^{+0.013}$ & $2.77 \pm 0.11$ \vspace{0.1cm} \\ 
& $0.3_{-0.008}^{+0.01}$ & $0.687 \pm 0.012$ & $2.8 \pm 0.1$ \vspace{0.1cm} \\ 
& $0.303_{-0.013}^{+0.015}$ & $0.684_{-0.015}^{+0.013}$ & $2.87_{-0.15}^{+0.18}$ \vspace{0.1cm} \\ 
& $0.299_{-0.008}^{+0.011}$ & $0.685_{-0.011}^{+0.013}$ & $2.88_{-0.11}^{+0.1}$ \vspace{0.1cm} \\ 
\hline
\multirow{3}{8em}{eBOSS} & $0.302_{-0.024}^{+0.037}$ & $0.663_{-0.026}^{+0.036}$ & $3.02_{-0.2}^{+0.21}$ \vspace{0.1cm} \\ 
& $0.33_{-0.034}^{+0.033}$ & $0.655 \pm 0.034$ & $3.24_{-0.19}^{+0.15}$ \vspace{0.1cm} \\ 
& $0.332_{-0.038}^{+0.04}$ & $0.667_{-0.031}^{+0.022}$ & $3.11_{-0.23}^{+0.22}$ \vspace{0.1cm} \\ 
& $0.3_{-0.017}^{+0.044}$ & $0.682_{-0.049}^{+0.021}$ & $3.29 \pm 0.16$ \vspace{0.1cm} \\ 
\hline
    \end{tabular}
    \caption{Peak posterior values and 68\% CIs for the cosmological parameters of interest resulting from the joint analyses discussed in Section \ref{subsec:joint_res}. All analyses conducted with $k_\mathrm{max.}=0.2\ h\ \mathrm{Mpc}^{-1}$ and the mariginalised likelihood. Each row is split into two. These correspond to results obtained with different analysis setups. From top to bottom, those are: sub-model $\mathcal{M}_1$ with the classic style prior, sub-model $\mathcal{M}_3$ with the classic prior, sub-model $\mathcal{M}_1$ with the Jeffreys prior, and model $\mathcal{M}_3$ with the Jeffreys prior.}
    \label{tab:combo_res}
\end{table}

%%%%%%%%%%%%%%%%%%%%%%%%%%%%%%%%%%%%%%%%%%%%%%%%%%

% Don't change these lines
\bsp	% typesetting comment
\label{lastpage}
\end{document}